\documentclass{aa}  
\usepackage{graphicx}

\usepackage{txfonts}
\usepackage{subfigure}
\usepackage[hidelinks,colorlinks=true,linkcolor=blue,citecolor=blue,urlcolor=black]{hyperref}
\usepackage{gensymb} %for \degree
\usepackage{txfonts}
\usepackage{array}
\usepackage{threeparttable}
\usepackage{booktabs}
\usepackage{makecell}
\usepackage{siunitx}
\usepackage{natbib}
\bibpunct{(}{)}{;}{a}{}{,} % to follow the A&A style

\begin{document}

 %  \title{Internal dynamics of the young star cluster Berkeley 59}
      \title{Internal dynamics and structure of Cepheus OB4}
      \subtitle{The asymmetric expansion of Berkeley~59}

   \author{B. Wiesneth\inst{1,2},
   K. Mu\v{z}i\'c\inst{2,3}, 
   V. Almendros-Abad\inst{4}
          }

 \institute{Fakultät für Physik und Erdsystemwissenschaften, Universität Leipzig, Linnéstraße 5, 04103 Leipzig, Germany            
         \and
Departamento de Física, Faculdade de Ciências, Universidade de Lisboa, Edifício C8, Campo Grande, 1749-016 Lisbon, Portugal
         \and
Instituto de Astrofísica e Ciências do Espaço, Faculdade de Ciências, Universidade de Lisboa, Ed. C8, Campo Grande, 1749-016 Lisbon, Portugal\\
\email{kmuzic@fc.ul.pt}
        \and
            Istituto Nazionale di Astrofisica (INAF) – Osservatorio Astronomico di Palermo, Piazza del Parlamento 1, 90134 Palermo, Italy\\
             }

   \date{Received; Accepted}

  \abstract
   {Accurate measurements of the internal dynamics of young stellar clusters can be used to extract crucial information about their formation process. With $Gaia$, we are now able to trace stellar motions and study the dynamics of star clusters with unprecedented precision. A fundamental requirement for this analysis is a well-defined and reliable list of probable members.}
   {In this work, we examine a 2$\degr$-radius region in Cepheus OB4, centered on the young cluster Berkeley 59, to build a reliable candidate member list. Our catalog enables the determination of structural and kinematic parameters of the cluster, as well as other properties of its stellar population.}
   {We compiled a catalog of optical and near-infrared photometry, along with precise positions and proper motions from $Gaia$ DR3, for sources in the Cepheus OB4 field.  Membership probabilities were determined using a probabilistic random forest algorithm and further refined by requiring HR diagram positions consistent with a young age. From a list of 1030 probable members, we estimate a distance of $1009\pm12\,$pc to Berkeley~59.
   Masses, extinction, and ages were derived by fitting the spectral energy distributions to atmospheric and evolutionary models, while internal dynamics was analyzed using proper motions relative to the cluster's mean motion.}
   {Berkeley~59 exhibits an asymmetric expansion pattern with velocity increasing outward and a preferred motion toward the north. The IMF between 0.4 and 7\,M$_\odot$ follows a single power law ($dN/dM\propto M^{-\alpha}$), with the slope $\alpha = 2.3 \pm 0.3$, consistent with Salpeter’s slope and previous studies in the region. The region’s median age, estimated from the HR diagram, is 2.9\,Myr. The velocity dispersion of Berkeley 59 exceeds the virial velocity dispersion derived from its total mass ($650\pm30$\,M$_\odot$) and half-mass radius ($1.71\pm 0.13$\,pc). The 2D motions of a stellar group located $\sim 1\degr$ north of Berkeley~59 provide further support for the previously proposed triggered star formation scenario.
   }
   {}
  

   \titlerunning{Internal dynamics of Cepheus OB4}
    \authorrunning{Wiesneth et al.}
   \maketitle
%
%-------------------------------------------------------------------
\section{Introduction}

Stars predominantly form in clusters and associations, emerging within giant molecular clouds through fragmentation and gravitational collapse \citep{lada&lada03,gutermuth09,wright2020}. These young stellar groups often retain the turbulent and substructured nature of their parent clouds, potentially reflecting large-scale dynamical processes \citep{tan2000,inutsuka15,sills18}.  
%such as galactic shear, cloud-cloud collisions, or stellar feedback-driven compression 
The subsequent evolution may be strongly influenced by the formation of massive stars, which expel the surrounding gas through intense radiation and stellar winds, thus reducing the gravitational binding of the system  \citep{goodwin2006}. 
%As stars evolve within these dense regions, their interactions and internal kinematics shape the structure of the forming cluster.
The extent to which embedded clusters and their substructures survive this phase and merge into bound stellar systems remains an open question, but kinematic signatures of expansion can provide key evidence of ongoing dispersal \citep{parker14,sills18, armstrong24}.

One of the primary challenges in studying the early phases of star cluster formation has been obtaining precise kinematic measurements of young stars. Data from the $Gaia$ space mission \citep{gaiamission:gaiacollab:2016} have revolutionized our ability to trace stellar motions and study the dynamics of star clusters with unprecedented precision \citep[e.g.][]{Kuhn:berk59_data_young_clusters_gaia, cantat-gaudin19,meingast21}. In particular, the expansion has been reported for a number of young clusters \citep{pang21,guilherme-garcia23,dellacroce24} and star-forming regions and OB associations \citep{zari19,armstrong22}. The analysis of \citet{dellacroce24} reveals that a significant fraction of clusters younger than 30 Myr appear to be expanding, while the older clusters are mostly consistent with an equilibrium configuration. However, the majority of these studies do not include very young populations (a few Myr), which are crucial for investigating the kinematic patterns linked to the early dynamical evolution of these systems. The younger clusters and star-forming regions are affected by (often differential) reddening, which complicates membership determination and typically requires a more individual approach. 
\citet{Kuhn:berk59_data_young_clusters_gaia} find that 75\% of the 28 clusters younger than 5 Myr exhibit expansion, with those still embedded in molecular clouds being less likely to be expanding than those that are partially or fully revealed. Additionally, they find no evidence of sub-group or multiple cluster mergers, suggesting that if hierarchical cluster assembly occurs, it must take place quickly during the embedded phase.
In $\lambda$~Ori, \citet{armstrong24} find evidence of a significant substructure, though this is preferentially located away from the central cluster
core, which is smooth and likely remains bound. They find strong evidence for expansion, which also appears asymmetric, with the maximum rate of expansion being
directed nearly parallel to the Galactic plane. 
In their 3D kinematic study of 18 star clusters and associations, \citet{wright19} and \citet{wright24} report that the vast majority ($\sim95\%$) of groups show expansion, which is also predominantly asymmetric.  
These findings challenge a simple residual gas expulsion models predicting a radial expansion
pattern. On the other hand, the observed expansion in NGC\,2244 \citep{lim21, Muzic-etal:random_forest_rosette} appears to be radially symmetric. 

In this work, we concentrate on the Cepheus OB4 region \citep{kun2008, rossano83}, which is dominated by the central cluster Berkeley~59. The age of the cluster is $\sim$2 Myr \citep{pandey2008, majaess08} and its distance $\sim$1.1
kpc \citep{Kuhn:berk59_data_young_clusters_gaia}. The cluster hosts several OB stars in its core \citep{skiff14}, and is situated within the surrounding H
II region Sharpless 171 (S 171). S 171 is expanding due to interaction with hot stars of Berkeley~59 \citep{gahm2022}. To the north of Berkeley~59, we find the nebula NGC 7822, labeled also as BRC2 (bright-rimmed cloud) by \citet{sugitani91}, which is an integral part of S 171. Previous studies investigated the cluster at optical bands \citep{pandey2008, eswaraiah12,Panwar:Berkeley59_2018}, and by X-ray, near- and mid-infrared \citep{koenig12,rosvick13,Getman-etal:SFinCs,mintz21}. \citet{Kuhn:berk59_data_young_clusters_gaia} combine these methods with astrometric data from Gaia; the results relative to the expansion/contraction state of the cluster remain ambiguous due to large uncertainties.
Recently, \citet{Panwar2024} presented the first substellar initial mass function (IMF) in Berkeley~59.
In this paper, we perform a new membership analysis of Cepheus OB4, using the supervised machine learning algorithm Probabilistic Random Forest (PRF; \citealt{Reis:PRF}). The same method has previously been tested in the Rosette Nebula \citep{Muzic-etal:random_forest_rosette}, doubling the number of known members of the region, and allowing to detect a clear expansion pattern of the central cluster NGC\,2244 and the motions of other groups in its vicinity. 

This paper is structured as follows. In Section~\ref{sec:datasets}, we present the details of the dataset used in this work, including photometric and astrometric catalogs. The primary membership selection method using the PRF algorithm is given in Section~\ref{sec:PRF}, followed by the additional constraints on the membership in Section~\ref{sec:additional_membership}.
We present the
analysis of spatial and kinematic properties of the selected members (Section~\ref{sec:kinematics}).
Section~\ref{sec:be59} is centered on the properties of Berkeley~59, including its mass function, and a discussion of its dynamical state. Finally, Section~\ref{sec:summary} summarises the work, and lists the main conclusions.  
   
\section{Dataset}
\label{sec:datasets}
\subsection{Catalog assembly}
\label{sec:main_cat_assembly}

The primary catalog is based on Gaia DR3 \citep{gaiamission:gaiacollab:2016, gaiad3:gaiacollab:2022}, which offers high-precision astrometric data along with optical photometry. A cone search with $2 \degree$ radius about the center of Berkeley 59, $(\alpha, \delta)$=($00:02:16.8, +67:26:28$) as resolved by SIMBAD \citep{simbad:wenger:2000}, yields about half a million sources. We use a $2 \degree$ radius to include other structures of the Cepheus OB4 association, which may be related to Berkeley 59 \citep{mintz21}. 
%By this the whole molecular cloud around the Berkeley 59 is contained in our analysis. 
Additional optical and near-infrared photometry was queried from 2MASS \citep{2masscollab:2mass_data} and Pan-STARRS data \citep{panstarrs:data_release}, in the same region on the sky. The catalogs are then cleaned to make sure only high-quality measurements are included. For the Gaia data, we only accept sources with a valid proper motion measurement. The 2MASS data was restricted by using the \emph{ph\_qual} flag, for which only values A, B or C were allowed. These quality flags ensure that the detection of a source is above a certain sensitivity level\footnote{See: \url{https://www.ipac.caltech.edu/2mass/releases/allsky/doc/sec1_6b.html##phqual}}. In the Pan-STARRS data, we disregard any measurements with quality flags 1, 2, 64, 128\footnote{See: \url{https://outerspace.stsci.edu/display/PANSTARRS/PS1+Object+Flags}}. The three catalogs were then joined in TopCat \citep{topcat:taylor:2005} using Gaia DR3 as the base catalog: we keep all the sources from Gaia and add to this additional information from the other two catalogs. This means that all the sources in the final catalog have proper motions, but may not have measurements in a complete set of photometric filters. This is an important point to consider when applying machine learning techniques, as will be discussed later in the text.
The matching radius used in TopCat was $1''$. 
The resulting catalog of the region after filtering contains 443,212 sources.

\subsection{Catalog completeness}
\label{sec:completeness}
\begin{table}
\caption{Photometric completeness intervals.}             
\label{tab:completeness}      
\centering                         
\begin{tabular}{c c c}       
\hline\hline                 
Filter & Bright end [mag] & Faint end [mag]\\   
\hline 
   G & 6.0 & 20.6\\
   g & 13.0 & 21.7\\     
   r & 13.0 & 20.8\\
   i & 13.0 & 19.9\\
   z & 13.0 & 19.1\\
   y & 12.5 & 18.6\\
   J & 8.0  & 16.5\\
   H & 6.0  & 15.7\\
   $K_S$ & 6.0 & 15.1\\
\hline                                  
\end{tabular}
\end{table}

\begin{figure}
   \centering
   \includegraphics[width=\hsize]{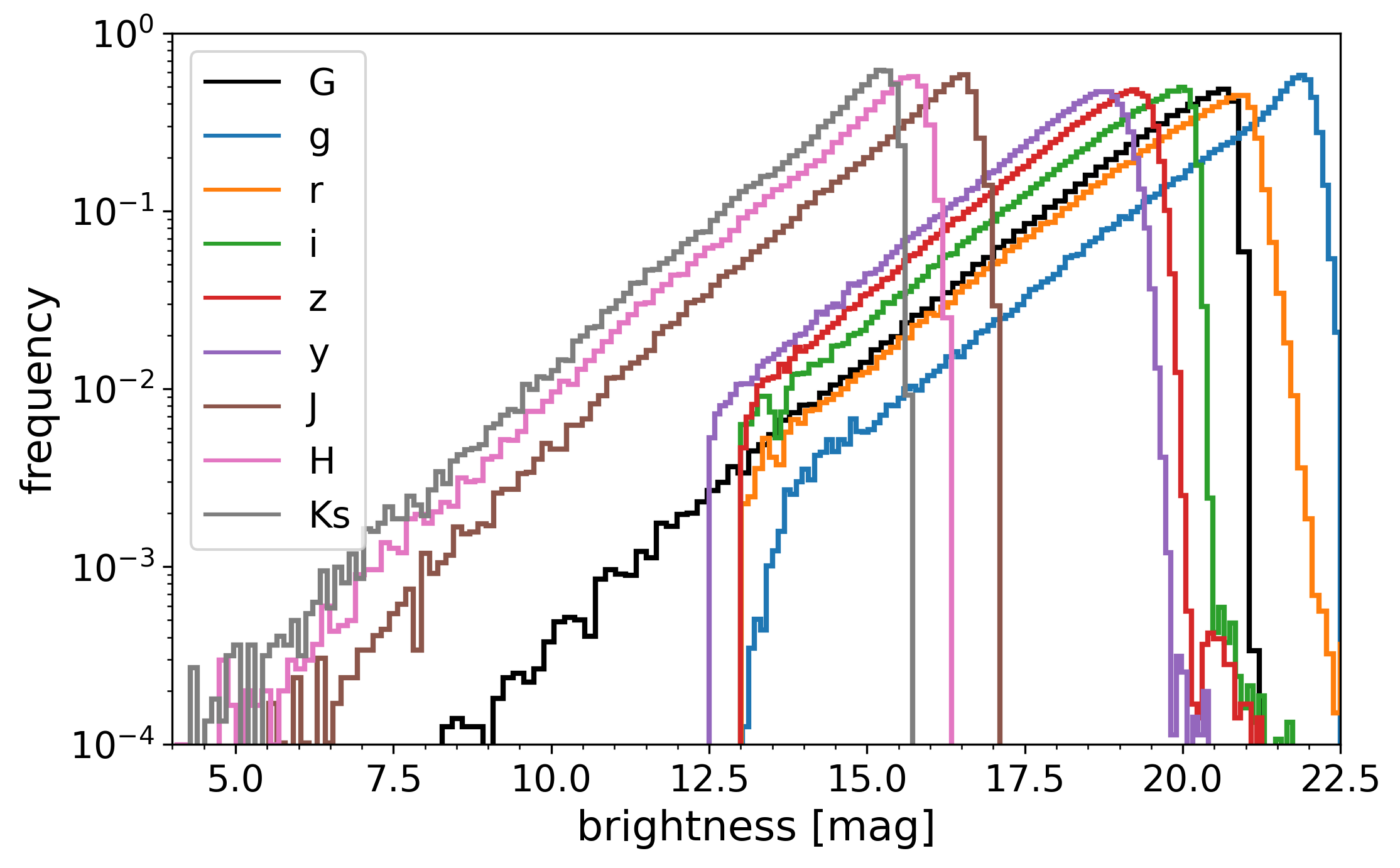}
      \caption{Density of sources in our final catalog as a function of magnitude.}
    \label{fig:completeness}
\end{figure}

The completeness limit of each photometric band was defined as the magnitude at which the density of sources reaches its maximum (Fig. \ref{fig:completeness}). In Table \ref{tab:completeness}, the completeness intervals are presented.  
The faint end is equivalent to masses between 0.1 and 0.6 M$_\odot$, depending on the filter, for the age of 2 Myr, distance of $\sim 1000$\,pc, and an average extinction of A$_V$=4\,mag \citep{Panwar:Berkeley59_2018}, according to BHAC15 models \citep{baraffe15}. Since the shallowest filter (in terms of mass; Pan-STARRS $g$) will not be used in the final run of the selection, our results are limited by the Gaia $G$- and Pan-STARRS $r$-bands, which is around 0.4M$_\odot$. This is the limit we adopt for the analysis later in the paper.

\section{Membership determination using the Probabilistic Random Forest Classifier}
\label{sec:PRF}

In this section, we describe the methodology for membership determination using the PRF algorithm. 
 In Section~\ref{sec:cons_train_set}, we describe in details the training set, while in Section~\ref{sec:prfmodel}, we present details of the model construction and its evaluation.

\subsection{Construction of the training set}
\label{sec:cons_train_set}
We are dealing with a classification problem with two classes: members of the cluster, and other sources which we call non-members. Thus, the training set consists of the lists of probable members and non-members, which are combined to form the final training set. In the following, we describe how these two lists have been created.
This step is critical since in supervised machine learning problems the quality of the model is defined by the quality of the training set. 
\subsubsection{Member list}
\begin{figure}
  \resizebox{\hsize}{!}{\includegraphics{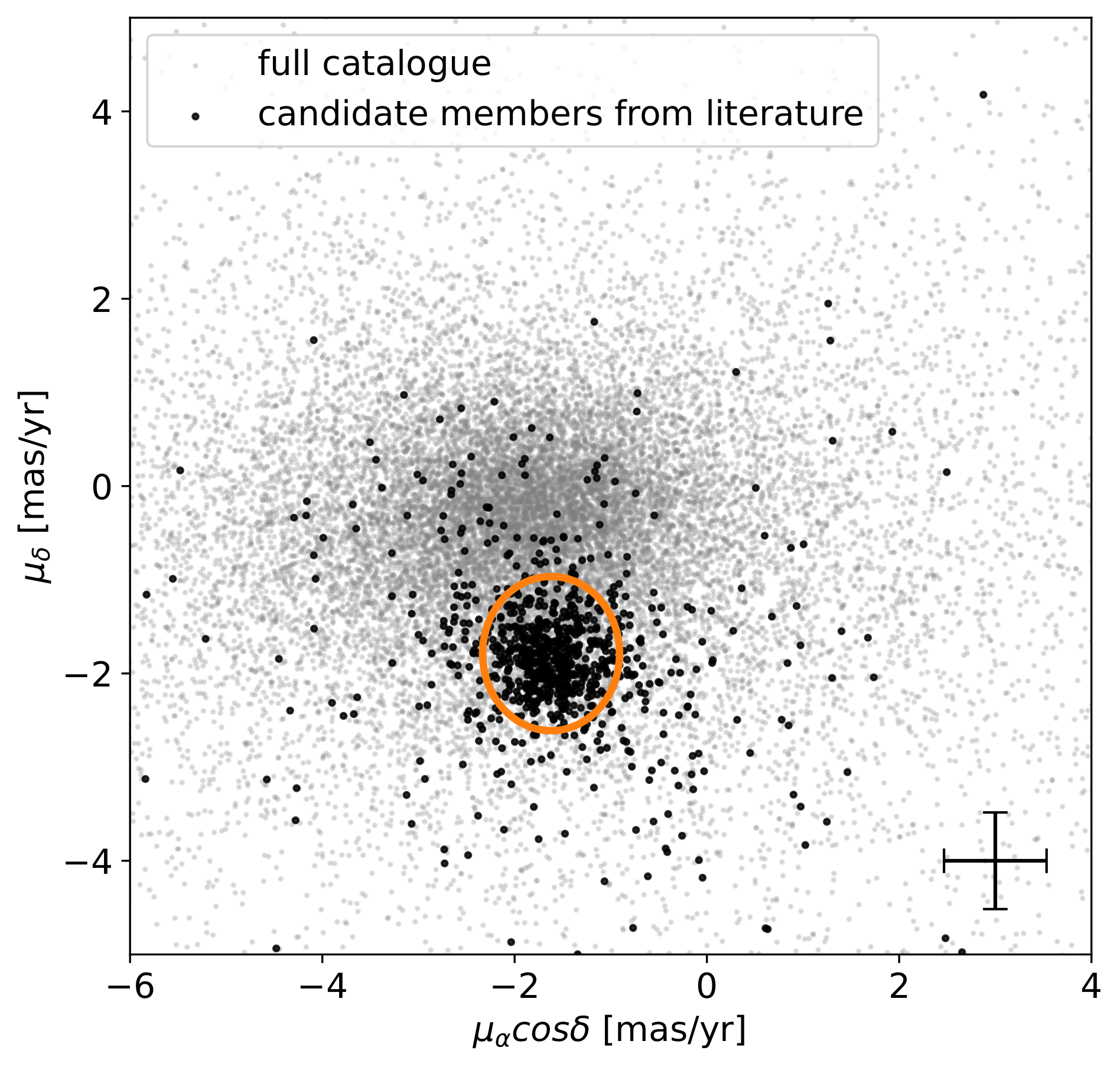}}
  \caption{Gaia DR3 proper motions of objects in Cepheus OB4 (grey dots). Black dots represent the 860 candidate members from the literature within a $30\arcmin$ radius of Berkeley 59 center. The orange ellipse indicates the proper motion selection criterion for constructing the member training set. In the lower right corner, we show the mean proper motion uncertainty.
  }
  \label{fig:pm}
\end{figure}
\begin{figure*}
   \centering
   \includegraphics[width=17cm]{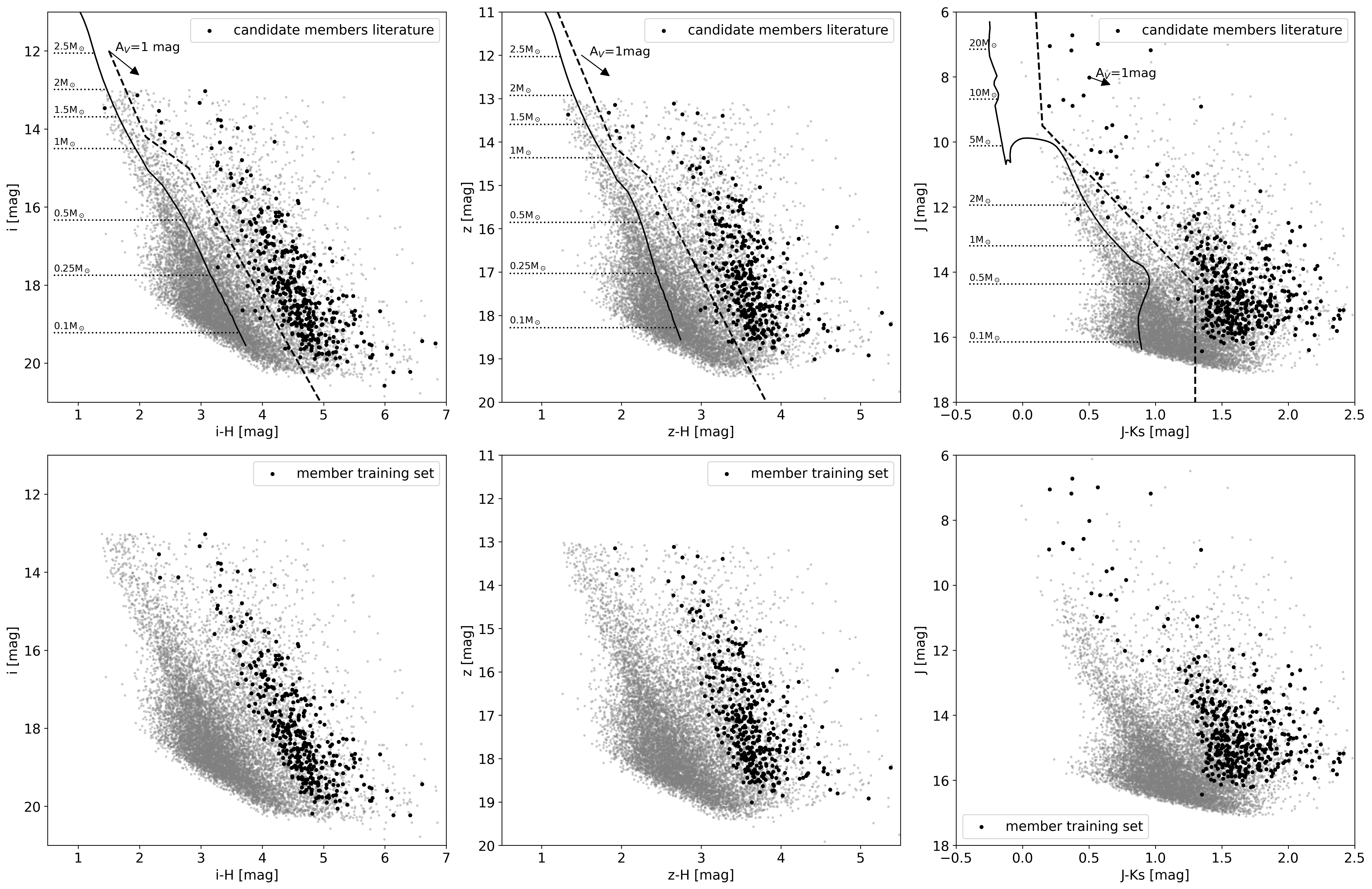}   
   \caption{Color-magnitude diagrams used for training set selection. Black dots in the top panels are candidate members from the literature, while the grey dots mark all the sources in the field. The dashed lines show the selection criteria, and the solid lines represent the 2 Myr PARSEC isochrones, shifted to a distance of 1100 pc \citep{Kuhn:berk59_data_young_clusters_gaia}. In the lower panels, the black dots represent the probable members selected for the training set, after the proper motion, color, and parallax filters.}
   \label{fig:cmd}
\end{figure*}
\begin{figure}
  \resizebox{\hsize}{!}{\includegraphics{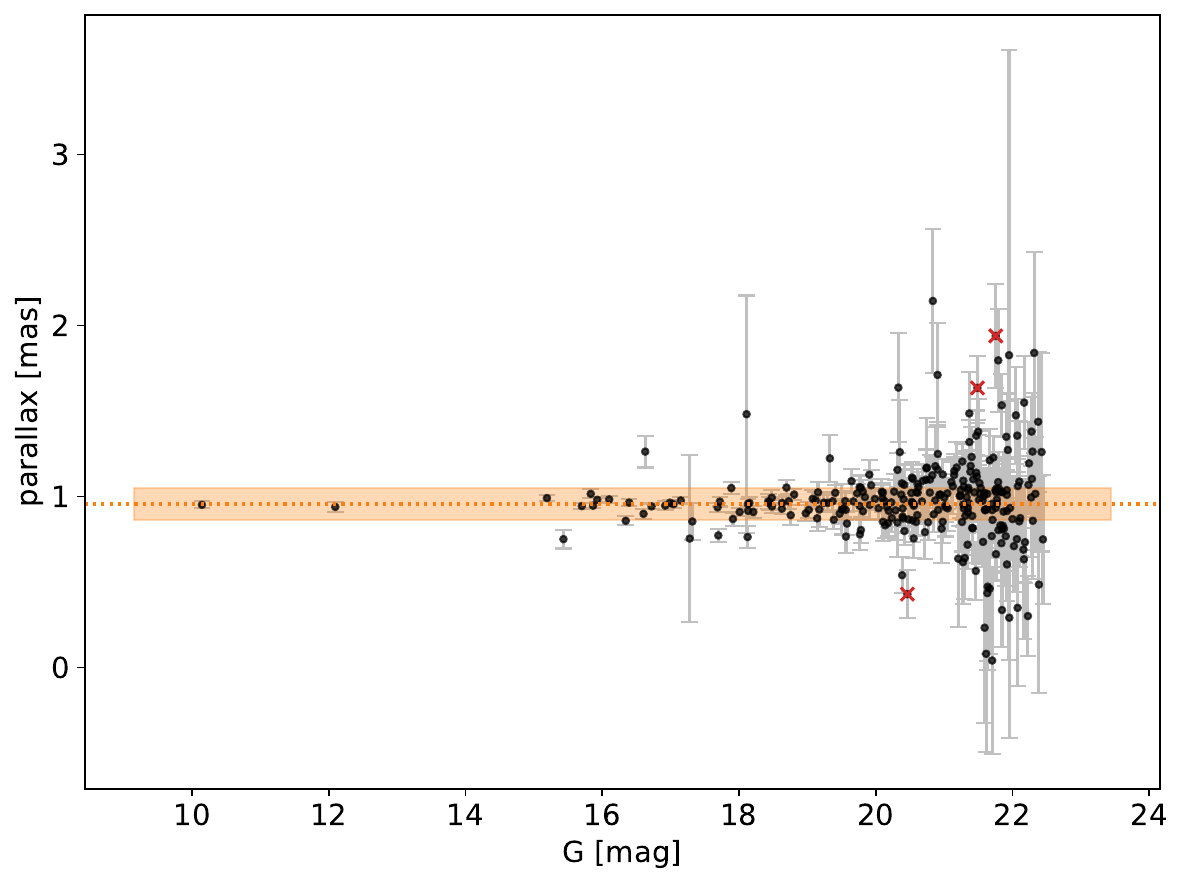}}
  \caption{Parallax measurements for the member training set candidates, selected based on proper motions and colors (black dots). The dotted orange line represents the weighted average of the parallaxes, while the shaded area marks $\pm1\sigma$ range. The red crosses mark the sources that have been excluded from the final member training set (see text for the details of the applied criterion).  }
  \label{fig:parallax_sel}
\end{figure}

To construct a high-probability member list, we combine the members catalogs from \citet{Getman-etal:SFinCs}, \citet{mintz21}, and \citet{Panwar:Berkeley59_2018}. 
The selection in the first two catalogs has been performed based on the X-ray and/or mid-infrared properties of the sources, while the third one used optical color-magnitude diagrams for the selection.
 We limit our selection of members to sources within $30 \arcmin$ of the center, where there is the highest concentration of the sources belonging to Berkeley 59. 
After cross-matching the list of previously reported members with our catalog, we are left with 860 common sources in this area. 

Next, we apply a cut to this member list based on the proper motions of the stars. 
Fig. \ref{fig:pm} shows the proper motions of the sources in a 30$'$ field around Berkeley 59 (grey dots), along with the probable members from the literature (black dots). The mean proper motion (3-$\sigma$-clipped) of the sources show as black dots is $\mu_{\alpha}\cos{\delta} = -1.62 \pm 0.59\, \mathrm{mas \, yr^{-1}}$ and $\mu_{\delta} = -1.79 \pm 0.69\, \mathrm{mas \, yr^{-1}}$. The orange ellipse has the semi-major axes equivalent to $1.2\sigma$ in $\alpha$ and $\delta$ and is centered at the mean proper motion. The 500 sources inside this ellipse were retained in the training set.  
The extent of the ellipse is somewhat arbitrary and is chosen with the intention to be conservative, rather than complete. 

This list can be further refined by using the position of the stars in a color-magnitude-diagram (CMD; Fig. \ref{fig:cmd}). Young stars are expected to have redder colors and therefore to appear to the right of the main bulk of stars. The three colors used are determined visually by comparing various diagrams in TopCat \citep{topcat:taylor:2005} and observing which ones give the clearest visual separation between members (redder colors) and non-members (bluer colors). The dashed lines mark our selection criteria, roughly aided by the shape of the 2 Myr PARSEC isochrone \citep{Bressan-etal:parsec_isochrone,Pastorelli-etal:parsec_isochrone}, which is depicted as the black solid line and has been shifted to a distance of 1100 pc. The sources (black dots) located to the right of the dashed lines in all three CMDs were retained in the member training sample. Stars that are missing some of the photometric points are retained if they pass the cuts in the remaining diagrams.
This cut reduces the member list to 468 sources. %Stars that are missing some photometry bands are still selected if they pass the selection criterion in at least one of the CMDs. By this we don't introduce a bias into our training set by rejecting sources just because they do not have a valid photometric measurement. The PRF can deal with those missing values. 

A final cut was applied using parallaxes. We reject any source located at a distance discrepant with the cluster distance using the parallax measurement of Gaia. All sources with the score $\zeta = |\varpi - \overline{\varpi}| / \sqrt{\sigma_{\varpi}^2 +\sigma^2}>3$ are removed from the member list, where $\varpi$ and $\sigma_{\varpi}$ represent the parallax measurement and uncertainty for each star, while $\overline{\varpi}$ and $\sigma$ are the weighted mean and standard deviation of the entire sample, correspondingly. The 6 excluded sources are marked with red crosses in Fig. \ref{fig:parallax_sel}. We are left with 462 probable member sources for the final training set.

\subsubsection{Non-member list}
The non-member list is sampled from the annulus with inner radius $1 \degree$ and outer radius $2 \degree$ from the center. This region is selected to avoid the densest part of Berkeley 59 and thus minimize the inclusion of potential members while staying close enough to sample similar extinction properties. We use two criteria to construct the non-memebers lists:
(1) sources whose individual proper motion do not intersect the orange ellipse in Fig. \ref{fig:pm}  within $1\sigma$; and 
(2) blue sources located to the left of the dashed selection lines shown in Fig. \ref{fig:cmd}. The final non-member list obtained by this procedure contains 282,714 unique sources.

\subsection{Training and evaluation of the model}
\label{sec:prfmodel}

In this section, we describe in detail the parameters used to construct the model, and present the results of its evaluation.

\subsubsection{Characteristics of the PRF}

There are several important points to discuss before constructing a PRF model. These are listed here.
\label{sec:character_prf}

During the construction of the training set, we saw that the main distinguishing features are their proper motions, their positions in CMDs and parallax measurements. Therefore, the main features that we use are the proper motions, parallaxes, magnitudes and colors constructed from the differences of magnitude pairs. In total, there are 39 available features. Generally speaking, it is desirable to perform a feature selection in order to reduce computation time, and potentially improve the performance of the model by avoiding overfitting. To this end, we use the recursive feature elimination procedure, in which an initial run of the model (as described in Section~\ref{sec:ev_prf}) is performed to extract the feature importance. Next, the model is run again after removing the least relevant feature, and again after removing the two least relevant features. This is repeated until arriving to some minimum number of features, which we set to 5. At each step, the training set is split 10 times in the same way as described in Section~\ref{sec:ev_prf}. From this, we calculate an average performance metric in the form of Matthews correlation coefficient \citep{mcc:matthew}. The performance of the model steeply increases until reaching the maximum at 16 features. After that, it keeps slightly decreasing, with some oscillations. We therefore keep the 16 most important features, to which we add the $J-K_S$ color. The reason for including this color specifically is the importance of the infrared data at the bright end, where the optical filters are largely saturated. Since the number of objects in this region is significantly smaller than at the faint end, infrared colors are deemed less important by the model, but after inspecting the resulting infrared CMDs, we decided to add this feature manually.

In the PRF algorithm, there exist various adjustable hyperparameters, yet their impact on the results is found to be insignificant. To assess this, we alter individual parameters while maintaining others unchanged at their default settings. We vary the parameter \emph{n\_estimators} (number of trees) from 1 to 1000, \emph{max\_depth} (representing the depth of each tree in the forest) from 1 to 50, and \emph{max\_features} (maximum number of features considered at each node split) from 3 to 17 (all possible features). We observe that the accuracies vary by no more than 1\% from the mean value across the tested range. As a result, we choose to adopt the default parameter values (\emph{n\_estimators}=100, \emph{max\_depth}=10, \emph{max\_features}=$\sqrt{N}$, where N=17 is the total number of available features).

While it is a requirement for all objects in our catalog to possess proper motion measurements, some of them lack one or more photometric measurements. 
A notable advantage of the PRF algorithm lies in its ability to handle missing data without requiring any specific treatment of the data. This is due to the representation of feature measurements as PDFs. During both the training and prediction stages, an object with a missing value for a particular feature is assigned a probability of 0.5 to propagate to either the member or non-member tree nodes. 

To create the training set, we combined the member list, consisting of 462 sources, with the non-member catalog, containing 282,714 sources. To compensate for this imbalance we resample the two classes, either oversampling the member class, undersampling the non-member class or a mixture of both. For this task, we use the Python package \emph{imbalanced-learn} \citep{Lemaitre-NAC:imbalanced_learn}. The \emph{sampling\_strategy} keyword was used to control the extent of random over- or undersampling. For instance, setting \emph{sampling\_strategy}=0.5 results in either undersampling the majority class or oversampling the minority class such that we obtain a ratio of 1:2 between minority and majority class. The combinations used are detailed in Table \ref{tab:prf_sampling}. The selection of sampling parameters aimed to achieve a reasonably balanced training set, i.e. keep the numbers in the same order of magnitude.

\subsubsection{Evaluation and comparison of PRF runs}
\label{sec:ev_prf}

The PRF was run six times (runs name A to F in Table~\ref{tab:prf_sampling}) with different sampling strategies and the same default hyperparameters.
We evaluate the performance of these runs using cross-validation. The main training data is split into a set of test and train data with a ratio of 1:4, using stratified sampling in order to preserve the ratio between the two classes. For each classifier, 50 test and train split samples are created randomly and the PRF is re-run for each of them. 

We use four different metrics to compare the outcome of the various runs. These are the F1 score, the area under the receiver operating characteristic curve (ROC\_AUC), the area under the precision recall curve (PR\_AUC), and the Matthews correlation coefficient. The results are given in Table~\ref{tab:prf_sampling} and shown in Fig.~\ref{fig:prf_scores}. The error bars were calculated as the standard deviation of the resulting scores from 50 random split samples.

As can be seen from Fig.~\ref{fig:prf_scores}, run C (the one with the smallest training set) has the worst performance in all four metrics. The remaining five runs performed similarly well (above 99\% in all metrics). We maintain those 5 runs and use them for classification on the full catalog. Each run assigns a certain membership probability to each source: the final probability was calculated as the arithmetic mean of the five values. 

In Fig. \ref{fig:feat_imp} we show the relative feature importance for the 17 used features in run F. The distribution looks similar for the other runs. We see that proper motion in Dec is by far the most significant feature as the cluster's proper motion differs the most in this direction from the field population. Next follow two optical-NIR colors, then proper motion in RA and in fifth place parallax. The cluster's proper motion in the RA direction is relatively similar to the field motion, so this feature is slightly less informative than proper motion in Dec. 
The strong performance of the classifier is primarily due to the careful construction of the training set, where we applied informed cuts in proper motion, colors, and parallax (Figs.~\ref{fig:pm}, \ref{fig:cmd}, \ref{fig:parallax_sel}), allowing a clear separation between the classes. 
Moreover, the deliberate selection of the 17 most informative features further enhanced performance.

\section{Further constraints on the membership}
\label{sec:additional_membership}
\subsection{High-probability candidates from the PRF}
\label{sec:highprobcands}

\begin{figure}
  \resizebox{\hsize}{!}{\includegraphics{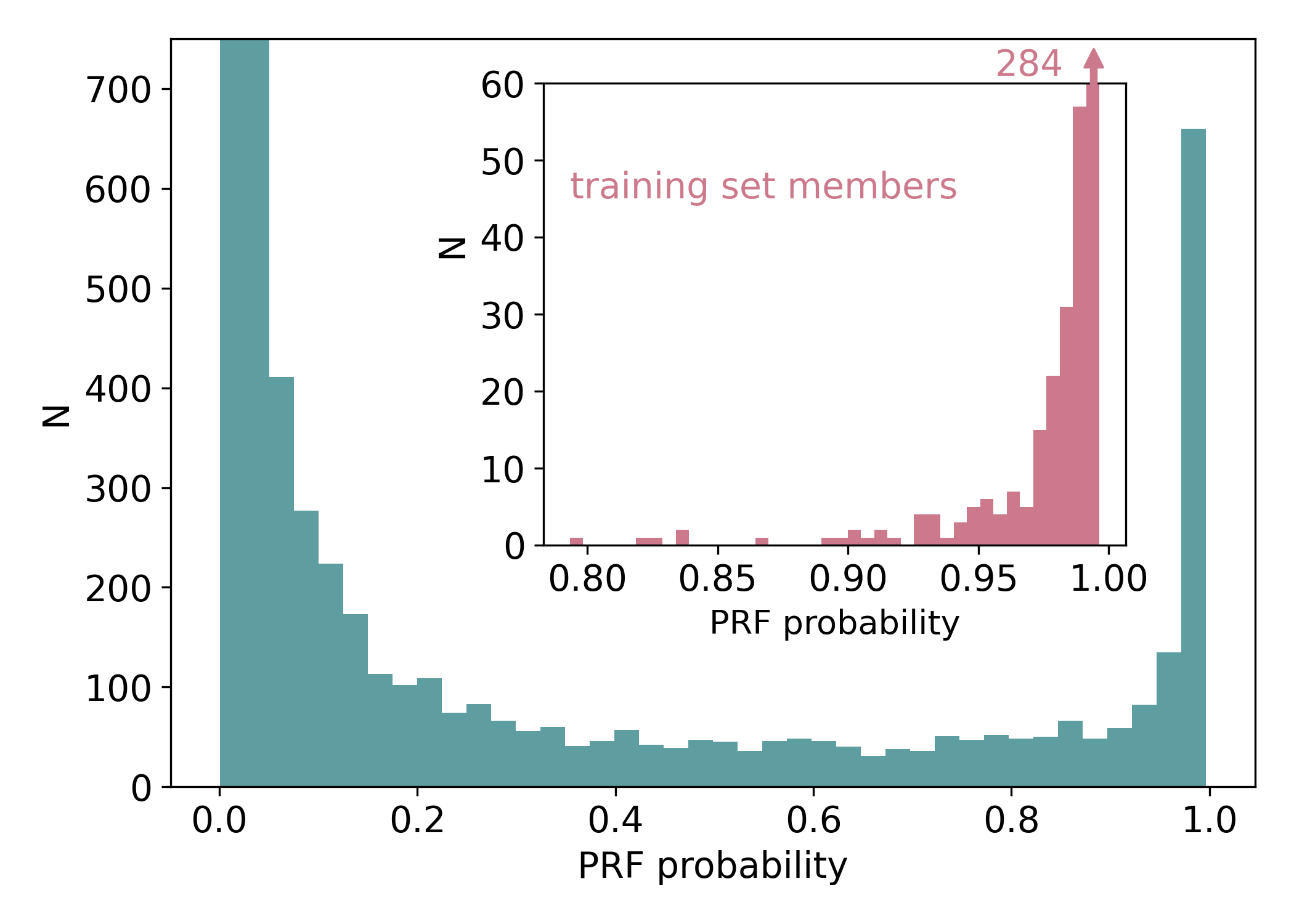}}
  \caption{Distribution of the membership probability of the final catalog limited to objects $30 \arcmin$ from the center for clarity. The four leftmost columns are capped for better readability. The inset diagram shows the prediction for all objects labelled as members in the training set. The rightmost column of the inset diagram is capped and its value depicted next to it.}
  \label{fig:prf_probs}
\end{figure}

The outcome of the procedure described in the previous section is a membership probability for each source. The green histogram in Fig.~\ref{fig:prf_probs} shows a distribution of these probabilities for the inner 30$'$ from the cluster center (the limit has been set for plotting clarity), while the inset diagram shows the prediction for all the objects identified as members in the training set. 454 out of 462 (98$\%$) are recovered with a probability of above 90\%, and all of them with a probability above 79\%. Considering the high recovery rate and aiming for a cleaner classification, we take objects with a probability $\geq 90\%$ to be members of the region. This results in a total number of 1224 candidate members. In the following, we further refine this selection.

To estimate the contamination rate in our sample, we apply the same PRF classifier described in Section~\ref{sec:ev_prf} to a nearby control field unrelated to Berkeley 59 and centered at $(\alpha, \delta$)=(23:44:20.5, +71:10:07), with a radius of $1^\circ$. The catalog for this control field was constructed using the same procedure outlined in Section~\ref{sec:main_cat_assembly}, and contains $\sim$127,000 objects. None of these objects was classified as a member with a probability higher than $80\%$, and only 47 ($\sim 0.04\%$) were assigned a probability of $50\%$ or higher. These findings suggest that the contamination rate in our sample is likely to be very low.

\subsection{SED fitting and HR diagram}
\label{sec:sed_hr}
Using the optical and infrared photometry, we construct spectral energy distributions (SED) for the PRF-selected candidate members and use the stellar synthetic atmosphere models to derive the effective temperature ($T_{\mathrm{eff}}$), extinction ($A_{\mathrm{V}}$), and surface gravity (log($g$)). 
For objects that show excess emission in WISE photometry, SED fitting is carried out using only the optical and near-infrared portions of the spectrum. Otherwise, the full available wavelength range is included in the fit. 
The metallicity is set to the Solar value. We use BT-Settl models \citep{allard11} and explore $T_{\mathrm{eff}}$ between 2000\,K and 20000\,K in 100 K increments. The $A_{\mathrm{V}}$ was varied between 0 and 10 mag in steps of 0.25\,mag, and log($g$) between 3.5 to 5.0, which is appropriate for young low-mass stars as well as field stars\footnote{As noted before in e.g., \citet{bayo17, Muzic-etal:random_forest_rosette}, the SED fitting process is not highly sensitive to log($g$), resulting in flat probability density distributions across the examined range.}. 
To obtain the best-fitting model, we perform a least square minimization, using the square of the photometric uncertainties as weights. Prior to this, we modified the uncertainties to 5$\%$ of the flux value in cases where the catalog values were smaller than that. Otherwise, we noted a systematic bias in the majority of our fits, where the best-fit model provided a consistently poor fit to the infrared photometry.
The model SEDs were scaled to match the median object fluxes.

\begin{figure*}
   \centering
   \includegraphics[width=\textwidth]{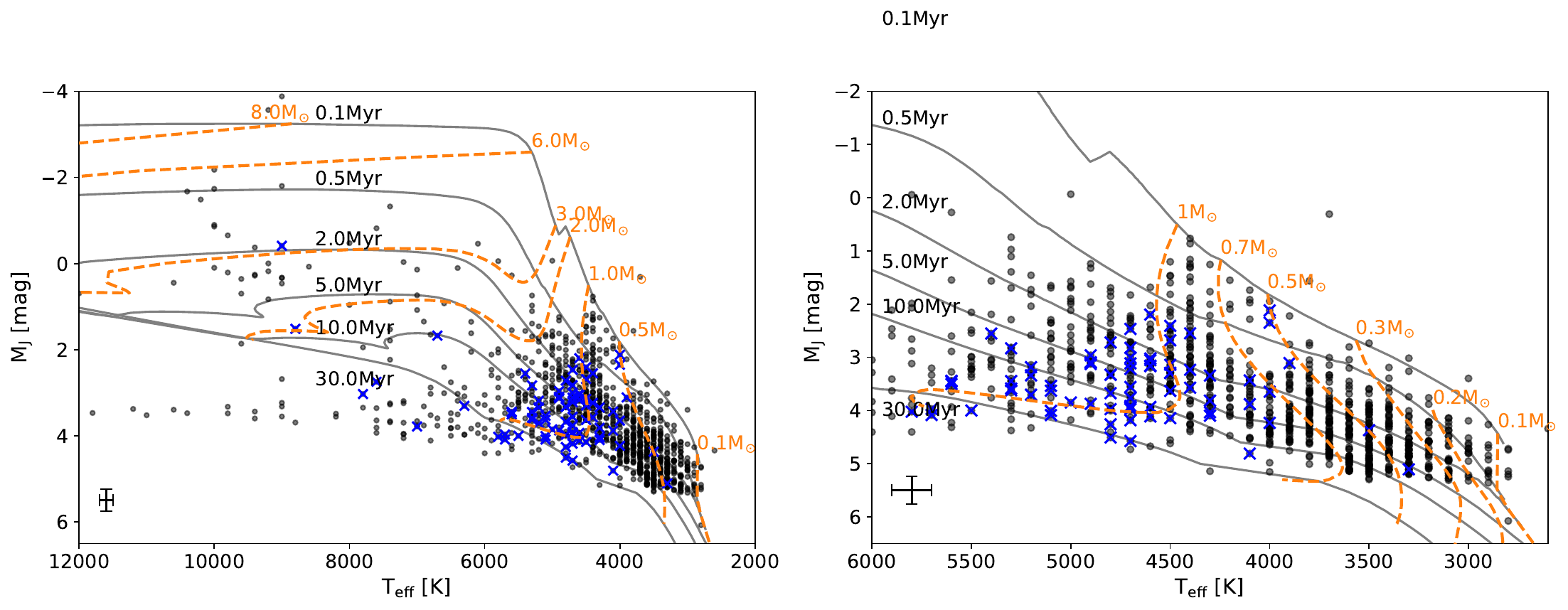}
      \caption{Hertzsprung-Russell diagram showing the candidates with membership probabilities $\geq$ 90$\%$ (black dots). The blue crosses mark those located in the southern filaments (see Sec.~\ref{sec:distance}). The isochrones (grey solid lines) and the lines of constant mass (dashed orange lines) are from the PARSEC series. A typical error bar is shown in the lower left corner. Objects located below the 30 Myr isochrone are considered contaminants. The right panel is the zoomed-in version of the one to the left.}
         \label{fig:hrd}
\end{figure*}

In Fig.~\ref{fig:hrd}, we show the Hertzsprung-Russell (HR) diagram constructed using the $T_{\mathrm{eff}}$ and $A_{\mathrm{V}}$ derived from the SED fitting, assuming the distance of 1009\,pc.  
The PARSEC evolutionary tracks \citep{Bressan-etal:parsec_isochrone, Pastorelli-etal:parsec_isochrone} for ages between 0.1 and 30 Myrs are shown as grey lines. The majority of the PRF-selected objects appear consistent with young ages, validating our selection method. However, there is a group of sources located below the 30\,Myr isochrone, which are likely background contaminants. These sources are all at the faint end of the apparent magnitude distribution ($G\gtrsim 19$) and have large parallax errors (mean $\varpi/\sigma_\varpi\approx 2$, compared to $\approx$9 for the full sample shown in the HRD). These sources comprise 10$\%$ of the PRF-selected member candidates, and were removed from the further analysis. At this stage, we are therefore left with 1106 member candidates.  

\subsection{Spatial distribution and distances}
\label{sec:distance}

\begin{table}
\begin{center}
\caption{Distances to various structures in the studied region} \label{tab:distance}
\begin{tabular}{lcc}\hline\hline
Region     &  d$_1$ [pc]\tablefootmark{a} & d$_2$ [pc]\tablefootmark{b}\\
\hline
all          & $1049 \pm 11$ & $1023 \pm 10$\\
all excl. south\tablefootmark{c}         & $1038 \pm 10$ & $1012 \pm 10$\\
Berkeley~59  & $1037 \pm 12$ & $1009 \pm 12$  \\
northern clump  & $1086 \pm 91$ & $1063 \pm 92$  \\
southern filaments  & $1817 \pm 188$ & $1794 \pm 188$  \\
\hline
\multicolumn{3}{l}{$^{a}$\footnotesize{from $Gaia$ EDR3 parallaxes}} \\
\multicolumn{3}{l}{$^{b}$\footnotesize{from $Gaia$ EDR3 parallaxes, after correcting }} \\
\multicolumn{3}{l}{\footnotesize{the parallax bias}}\\
\multicolumn{3}{l}{$^{c}$\footnotesize{excluding the sources in the southern filaments}}
\end{tabular}    
\end{center}
\end{table}
\begin{figure*}
\begin{subfigure}
   \centering
   \includegraphics[width=0.5\textwidth]{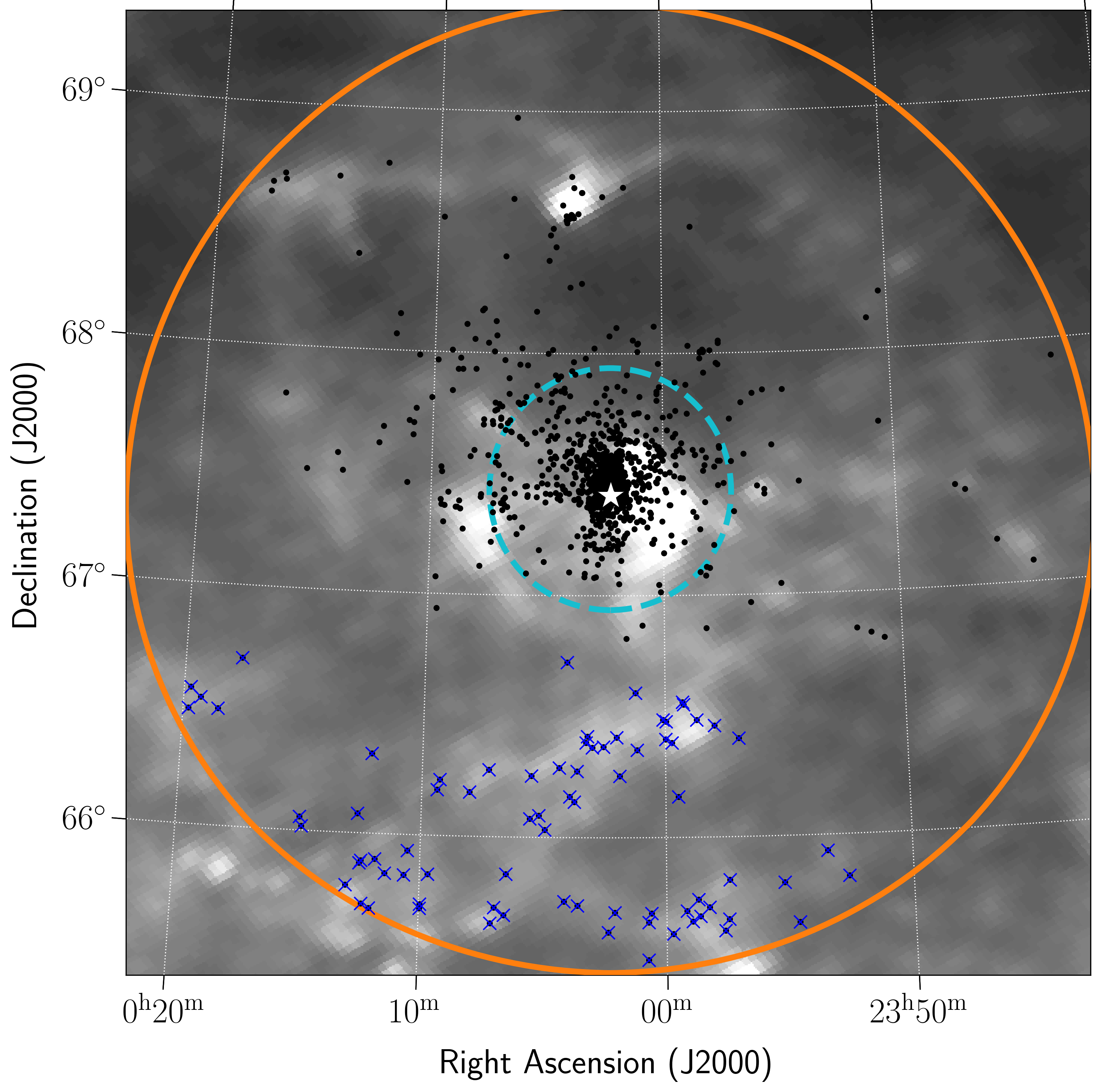}  
\end{subfigure} 
\begin{subfigure}
   \centering
   \includegraphics[width=0.5\textwidth]{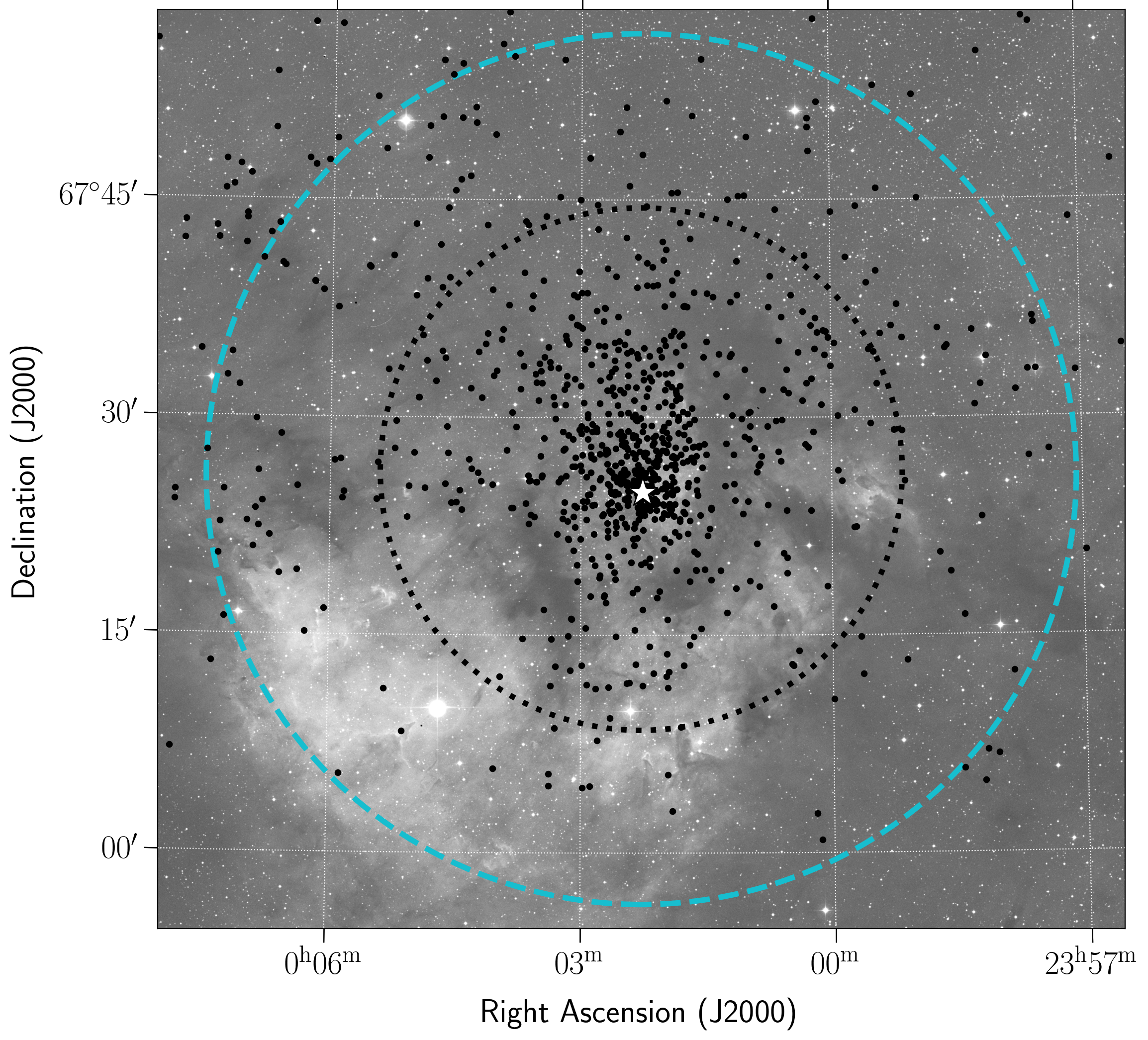}  
\end{subfigure} 
   \caption{Sources with membership probabilities $\geq 90\%$ and consistent with ages $<30$ Myr (black dots) on top of the Planck 857 GHz image (left) and DSS red image (right) of the region around Berkeley 56. The sources likely located in the background, associated with the southern filaments are marked with blue crosses. The solid orange circle encompasses the entire studied region (r=$2\degree$), the blue dashed one the region where the training sample has been selected (r=$30'$), and the black dotted one an extent of the cluster as derived from the radial profile (r=$18'$). The white star shows the position of the center of Berkeley 59 derived in section \ref{sec:centre_rad_distr}.}
   \label{fig:onskyplot_zoom_rect}
\end{figure*}

In Fig.~\ref{fig:onskyplot_zoom_rect}, we show the high-probability candidate members from the PRF, with ages below 30 Myr (black dots and blue crosses) on top of a Planck 857 GHz image \citep{Planck2016}. A zoomed-in version of the plot is shown in the right panel, on top of an image from Digitized Sky Survey (DSS\footnote{\url{https://irsa.ipac.caltech.edu/data/DSS/overview.html}}).
The region is dominated by the central cluster Berkeley 59. At the northern edge of the left panel in Fig.~\ref{fig:onskyplot_zoom_rect}, at the border of the HII region, there is a small concentration of sources previously dubbed cluster \#0 in \citet{mintz21} and BRC2 in \citet{sugitani91, ogura2002} and associated with the nebula NGC~7822 \citep{lozinskaya87}. Outside the Berkeley 59 region,  we find sources in regions associated with the higher dust concentration. In the lower half of the left panel the sources follow the filamentary structure of the molecular cloud.

The distance to the cluster is recalculated using the Gaia DR3 parallax measurements and applying the maximum-likelihood procedure as in \citet{Muzic-etal:deep_look_rosette} and \citet{Cantat-etal:max_likelihood_distance}. 
Furthermore, it has been shown that Gaia parallaxes contain a bias discovered and characterized via quasar measurements. This bias depends in a non-trivial way on position, color, and magnitude \citep{Lindegren-etal:gaia_bias}. To estimate the value of the parallax bias, we use the Python package\footnote{\url{https://gitlab.com/icc-ub/public/gaiadr3\_zeropoint}} based on the prescription given in \citet{Lindegren-etal:gaia_bias}. 
The bias has been removed from all the sources for which it could be calculated ($\sim 65\%$ of the high-probability sources). 

The distances are given in Table~\ref{tab:distance}. The first row quotes the distance obtained using all the high-probability sources. We also calculated the distance to different sub-regions of our field-of-view, including the cluster Berkeley~59 (centered at $(\alpha, \delta$)=(00:02:16.0, +67:24:52) and within a radius of $18'$; see Section~\ref{sec:centre_rad_distr}), the clump in the northern part (cluster \#0 in \citealt{mintz21}), and to the sources associated with the two dust filaments to the south of Berkeley~59. 
While the northern clump appears to be at a similar distance as the cluster, the sources associated with the southern dust filaments seem to be more than 700 pc farther away. Despite having proper motions and colors similar to other cluster members, their parallaxes show significantly poorer agreement. Notably, the mean $\varpi/\sigma_\varpi$ for these sources is 2.4, indicating that their large parallax uncertainties allowed them to be classified as high-probability members. Interestingly, if we applied a distance modulus corresponding to ~1800 pc, the southern sources would shift upward by 1.3 mag in the HR diagram, placing them in the 0.1–5 Myr region. This suggests that while these sources are likely genuinely young, they are situated well beyond the Berkeley 59 region.
We exclude the 76 sources in the southern filaments ($\delta < 66.8\degree$; blue crosses in Fig.~\ref{fig:onskyplot_zoom_rect}) from the further analysis of this paper, given that they are probably located well behind Berkeley~59 and not associated with it.
We also recalculate the distance to the region (second row in Table~\ref{tab:distance}). 

The distance to Berkeley~59 obtained here is slightly lower than $1100 \pm 50$ pc from  \citet{Kuhn:berk59_data_young_clusters_gaia} and $\sim$1.1\,kpc from \citet{gahm2022}, and in line with $1.00\pm0.06$\,kpc found in \citet{Panwar2024}. For the remainder of this paper, we will adopt the bias-corrected distance of Berkeley~59 of 1009\,pc.

\subsection{Summary of the member selection}
\label{sec:membershipsummary}

We summarize the remaining sources in the final sample after applying various member selection cuts: 

\begin{itemize}
    \item sources with the PRF probability $\geq90\%$: 1224;
    \item after removing objects below the 30\,Myr isochrone: 1106;
    \item after removing objects in the south ($\delta < 66.8º$): 1030.
\end{itemize}

The following analysis is based on these final 1030 member candidates, listed in Table~\ref{tab:members}.

\begin{table}
\centering
\tiny 
\caption{Members selected in this work, along with the $T_{\mathrm{eff}}$ and $A_{\mathrm{V}}$ from the SED fitting.}
\label{tab:members}
\begin{tabular}{ccccc}\hline\hline
 GaiaDR3 & RA\_ICRS & DE\_ICRS & $T_{\mathrm{eff}}$ & $A_{\mathrm{V}}$ \\
 & & & (K)  & (mag)\\
\hline
			528574787530796544 & 00:00:00.55 & 67:33:32.0 & 5500.0 &  6.00 \\
			528598396970480640 & 00:00:01.94 & 67:35:30.3 & 5800.0 &  6.25 \\
			528565407326062336 & 00:00:03.70 & 67:18:21.3 & 4500.0 &  6.50 \\
			528598431330216192 & 00:00:03.89 & 67:35:54.9 & 4200.0 &  3.50 \\
			528598289592091520 & 00:00:04.67 & 67:34:18.1 & 3100.0 &  3.50 \\

\hline
\end{tabular}
        \tablefoot{The full table is available in electronic form at the CDS.
        }
\end{table}

\section{Spatial and kinematic properties of candidate members}
\label{sec:kinematics}

\subsection{Center and radial distribution of Berkeley 59}
\label{sec:centre_rad_distr}

\begin{figure}
  \resizebox{\hsize}{!}{\includegraphics{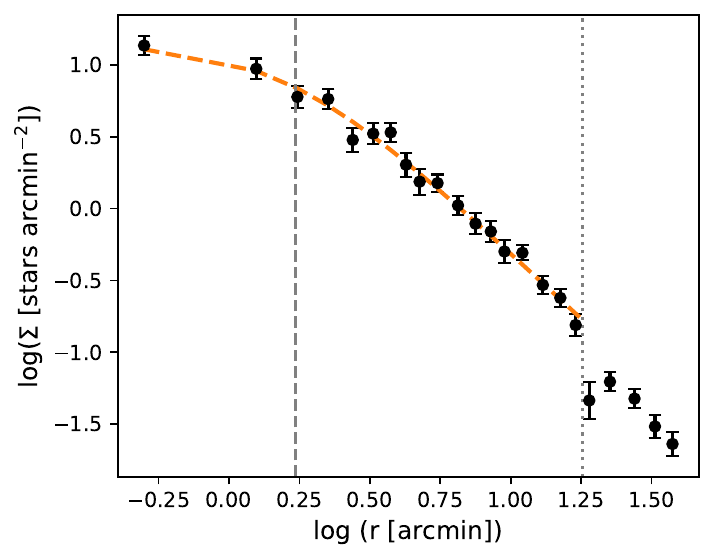}}
  \caption{Radial profile of Berkeley 59 in logarithmic scale. Each black dot shows the surface density of its corresponding annulus about the center. The dashed orange line is the best fit EFF profile \citep{Elson:radial_profile} up to a radius of $30'$(grey dotted line). The vertical dashed line represents the core radius derived from profile fitting, and the dotted one marks the approximate outer radius
of the cluster (r=18$'$).}
  \label{fig:radial_profile}
\end{figure}

We estimate the stellar surface density distribution of the cluster using a two-dimensional Gaussian kernel density estimator (KDE). The point of maximum density is taken as the center of the cluster, which is located at $(\alpha, \delta)$=(00:02:16.0, +67:24:52). 
The field is then divided in concentric annuli around the central point, and the radial profile (Fig. \ref{fig:radial_profile}) is derived by counting the stars inside each annulus and dividing by the respective area. 
At the radius of about $18'$, the density distribution has a sharp drop, which we take as an indication of reaching the
cluster's outer radius. This radius corresponds to $\sim$5.3 pc at a distance of Berkeley~59, and is marked with the vertical dotted line in Fig.~\ref{fig:radial_profile}.

Limiting the extent of the cluster to $r=18'$, we fit the radial distribution by a generalized radial profile known as the Elson-Fall-Freeman (EFF; \citealt{Elson:radial_profile}), in the following form: 
\begin{equation}
    \Sigma(r) = \Sigma_0\Big[1+\Big(\frac{r}{a}\Big)^2\Big]^{-\gamma/2},
\end{equation}
where $r$ stands for the projected distance from the center of the cluster, $\Sigma_0$ is the central surface density, and $a$ is a scale parameter. The core radius, r$_c$ of the King profile \citep{king66} is then given by
\begin{equation}
    r_c=a\,\Big(2^{2/\gamma} -1\Big)^{1/2}
.\end{equation}
The parameters of the best-fits profile (orange line in Fig.~\ref{fig:radial_profile}) are: $\Sigma_0 = 14.0 \pm 1.9$ stars arcmin\textsuperscript{-2}, $a=1.62 \pm 0.23'$ and $\gamma=1.83 \pm 0.09$. With this, we obtain the core radius $r_c =1.72\pm0.25' = 0.50 \pm 0.07$\,pc, depicted as the grey dashed line in Fig. \ref{fig:radial_profile}. The core radius is in agreement with that derived by \citet{pandey2008}, which ranges between $1'$ and $1.9'$, depending on the selection of the sources included in the calculation.  
The slope $\gamma$ is close to that of a modified Hubble model ($\gamma$=2; \citealt{binneytremaine}) and is similar to several other young clusters \citep{portegieszwart10,kuhn14,kuhn17,miret19}.
The mean surface density of the cluster, $\Sigma_{mean}$, is $\sim$\,3.0\,stars\,arcmin$^{-2}$, equivalent to 35\,stars\,pc$^{-2}$ at a distance of 1009 pc.

 In Table~\ref{tab:parameters}, we give a summary of various physical parameters of Berkeley 59, including those derived in this section.

\subsection{Internal kinematics}

\begin{figure*}
   \centering
   \includegraphics[width=15cm]{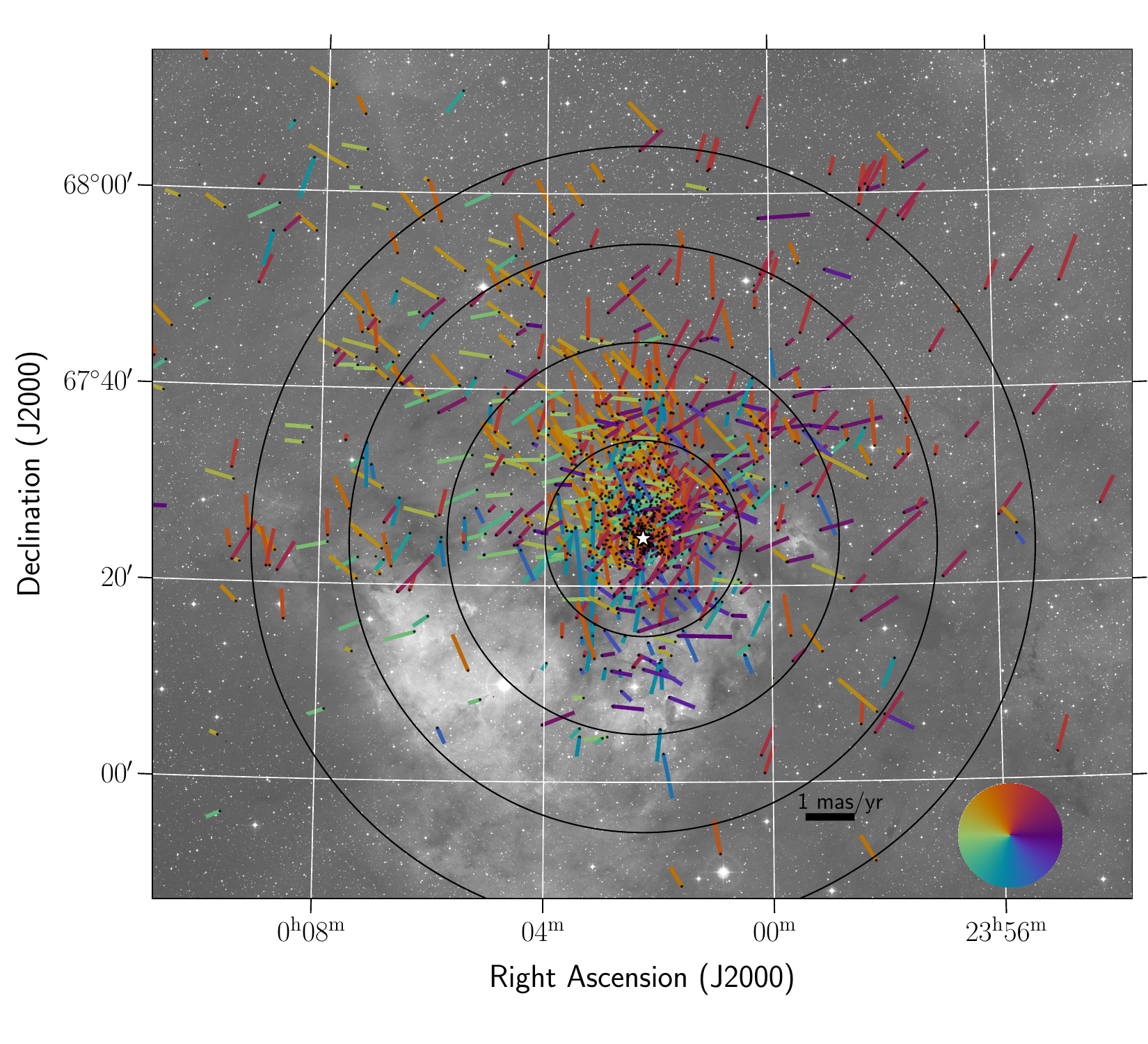}   
   \caption{Relative proper motion vectors of the stars in Berkeley 59 shown on top of a DSS red image. The color and direction of the bars extending from the black dots, that represent the sources, indicate the direction of the movement. We observe an expansion predominantly in the declination direction. The concentric circles mark radii of $(10 , 20, 30, 40)'$ from the center. The white star marks the point of highest surface density, taken as the center. For a zoomed-in version see Fig. \ref{fig:relpm_zoom}}
   \label{fig:relpm_nozoom}
\end{figure*}

\begin{figure*}
   \centering
   \includegraphics[width=15cm]{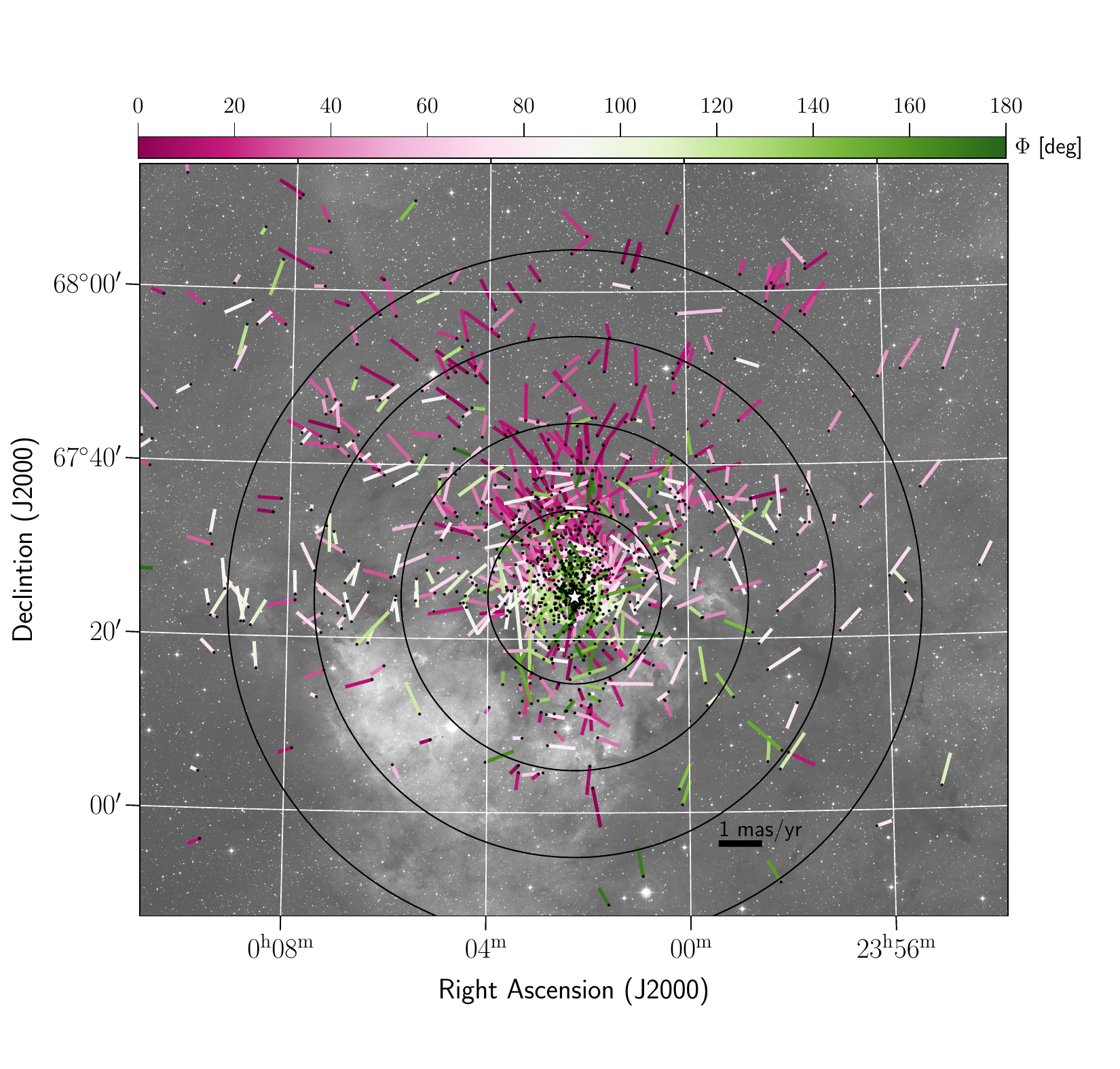}   
   \caption{Similar to Fig. \ref{fig:relpm_nozoom}. Here, the proper motion vectors are colored according to the angle between the line connecting the star to the center (white star) and its proper motion vector. Purple hues indicate a motion away from the center, and green towards the center. White-colored bars mark motion perpendicular to the line connecting the star to the center. Stars with white color may move in opposite directions with respect to each other.}
   \label{fig:pm_phi_nozoom}
\end{figure*}

\begin{figure}
  \resizebox{\hsize}{!}{\includegraphics{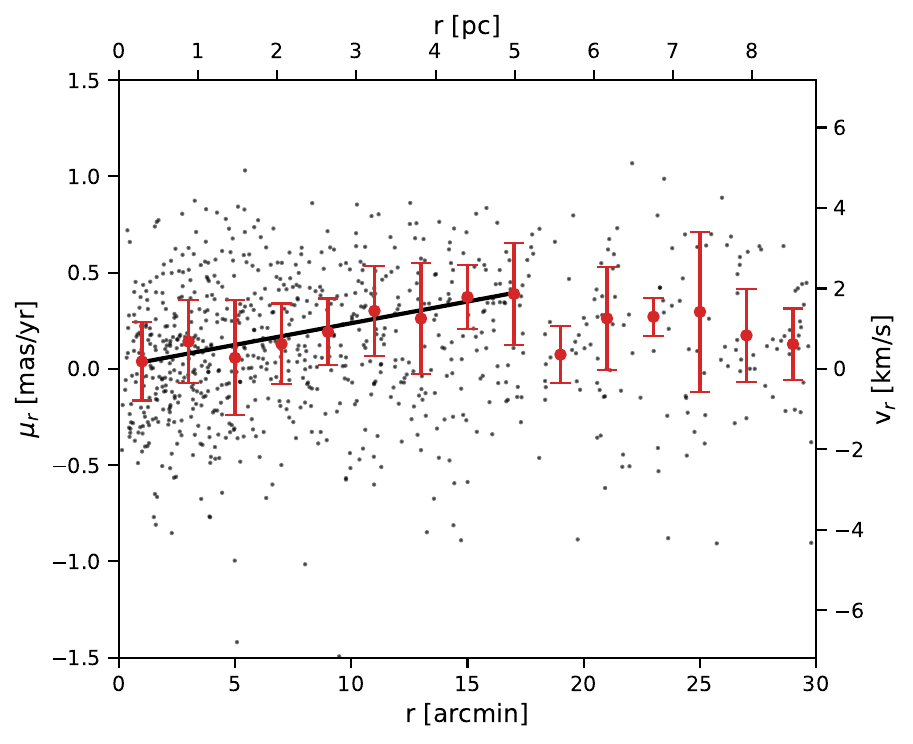}}
  \caption{Radial component of the relative proper motion ($\mu_r$) as a function of distance from the center of Berkeley 59. The red points and error bars show the values of $\mu_r$ in $2'$ bins, calculated as the weighted mean and standard deviation, respectively. A distance of 1009\,pc is assumed for the top and right axes. Positive values of $\mu_r$ indicate expansion. The black line shows a linear fit to the red points inside the radius of 18$'$, and has a slope of $0.022\pm0.003$\,mas\,yr$^{-1}$\, arcmin$^{-1}$ and an intercept of $0.013\pm0.028\,$mas\,yr$^{-1}$.}
  \label{fig:pmvsradius}
\end{figure}

\begin{figure}
  \resizebox{\hsize}{!}{\includegraphics{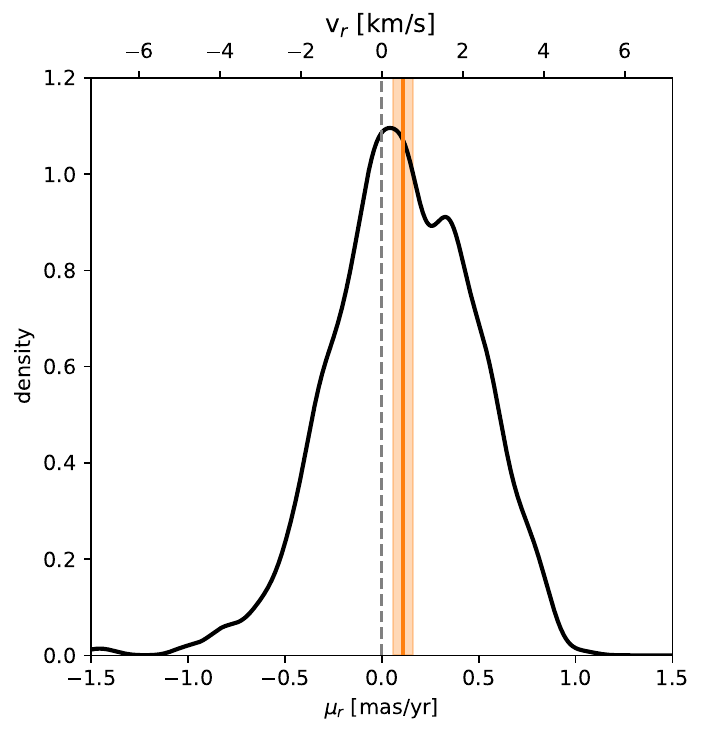}}
  \caption{Distribution of the radial component of the relative proper motion vector for Berkeley 59 ($ r \leq 18'$). The grey dashed line marks the velocity of zero, the solid orange vertical line represents the median of the distribution, and the orange shaded area spans a $3\sigma$ range around the median.
  }
  \label{fig:pm_radial_distr}
\end{figure}

To study the internal kinematics of the region, we start by calculating the error-weighted mean proper motion of Berkeley~59 within the radius of $18'$. The mean proper motion is then subtracted from all individual sources in the region. Next, we correct for the effect of perspective expansion (or contraction), which is caused by the projection of the cluster's radial velocity onto the proper motion depending on the position of a star relative to the projection center  \citep{vanLeeuwen:persp_correction,gaiacollab2018:persp_correction,Kuhn:berk59_data_young_clusters_gaia}. The radial velocity we use is v$_r$ = -13\,kms$^{-1}$, calculated as the weighted average of the numbers from \citet{conrad2017} and \citet{kharchenko2009}, who give R$_V$ = $-14.9 \pm 10.7$\,kms$^{-1}$ and $-12.5 \pm 7.0$\,kms$^{-1}$, respectively. As the center we use the position calculated in Section~\ref{sec:centre_rad_distr}. The effect of perspective expansion is on average small ($\sim$0.02 mas\,yr$^{-1}$), and more pronounced at the edges of the studied region ($\sim$0.15 mas\,yr$^{-1}$).

\subsubsection{Relative proper motions on the sky}

In Fig. \ref{fig:relpm_nozoom}, we show the relative proper motion of high-probability members, represented by black dots. In this plot, we show the central part of the studied region - the outer circle has a radius of $30'$.
A color-coded bar extends from each star to show the direction of its internal proper motion. 
There is no clear circular outward expansion pattern, which would reveal itself by resembling the color distribution similar to the disk shown in the lower right corner (see e.g. Fig. 12 in \citealt{Muzic-etal:random_forest_rosette}). However, to the north of the center, we can notice predominant orange-reddish hues indicating motion in the North-East direction, while to the south of the cluster center there are more blue-shaded lines, signaling a south-west motion. This indicates some level of expansion, roughly along the declination axis. To investigate this effect further, we introduce another color scheme that helps visualize expansion or contraction. To this end, we calculate the angle $\Phi$ defined as the angle between a star's relative proper motion and the line connecting it to the cluster center. $\Phi=0^{\circ}$ translates to the proper motion vector pointing away from the center, and $\Phi=180^{\circ}$ towards it. Fig. \ref{fig:pm_phi_nozoom} shows the same region on the sky as Fig.\ref{fig:relpm_nozoom}, but with the proper motion vector color-coded accordingly to the value of $\Phi$. Purple hues indicate an outward motion from the center, while green hues indicate an inward motion towards the center.
In this representation of the proper motions, the asymmetric expansion is seen more clearly, and in agreement with the interpretation of Fig. \ref{fig:relpm_nozoom}.
\citet{wright24} examined the 3D dynamics of 18 star
clusters and OB associations and found that the majority of the expanding groups do so asymmetrically. They suggest that, in the absence of external forces influencing the stars' motion, this asymmetry likely arises from non-spherical initial conditions or anisotropic velocity dispersions prior to expansion. 

In Fig.~\ref{fig:relpm_zoomnorth}, we zoom into the pillar to the north of the main cluster (cluster $\#0$ in \citet{mintz21}. As we have shown in Section~\ref{sec:distance}, the sources in this clump share the distance with Berkeley~59. We see that the sources in the clump predominantly move away from the cluster. Moreover, \citet{mintz21} find that the cluster $\#0$ contains a higher proportion of Class I sources relative to Class IIs than the central cluster Berkeley\,59. These two findings speaks in favor of the triggered star formation scenario, previously suggested by \citet{mintz21}.
Inspecting the number ratio between Class I and Class II sources as a function of the distance from Berkeley~59 center, \citet{mintz21} find that the ratio remains roughly constant out to $\sim 1\degr$, showing a steep increase further away from the cluster. Taking the ClassI/ClassII ratio as a proxy for age, the authors suggest that the sources closer to Berkeley 59 (basically all sources visible in our Figs.~\ref{fig:relpm_nozoom} and \ref{fig:pm_phi_nozoom}) were once likely cluster members. Their predominant outward motions speaks in favor of this hypothesis.    

\subsubsection{Radial component of the relative proper motion vector}

Next, we estimate the behavior of the radial component of the relative proper motion vector ($\mu_r$), defined as a projection of the relative proper motion vector onto the vector connecting the cluster's center with the star in question. Positive values of $\mu_r$ indicate a star's radial component pointing away from the center. In Fig.~\ref{fig:pmvsradius}, we show $\mu_r$ as a function of distance from the cluster's center. Sources are binned by radial distance with a bin size of $2'$. The weighted mean and corresponding standard deviation of $\mu_r$ in each bin are shown in red. We observe an increase of $\mu_r$ as we move outward from the center, up to a radius of about $20'$. After that, the mean $\mu_r$ drops and remains roughly constant up to $30'$ radius. The change of behavior occurs at a similar distance from the cluster center where the radial profile (Fig.~\ref{fig:radial_profile}) also shows a discontinuity.
 The increase of $\mu_r$ suggests a radially dependent expansion velocity with more stars moving faster outwards the further they are from the center. We fit a line to the observed increasing trend and obtain a slope $0.022\pm0.003$\,mas\,yr$^{-1}$\,arcmin$^{-1}$ =$0.37\pm 0.04$\,km\,s$^{-1}$\,pc$^{-1}$. This is significantly higher than $0.1\pm0.4$\, km\,s$^{-1}$\,pc$^{-1}$ previously found in \citet{Kuhn:berk59_data_young_clusters_gaia}, who performed a similar analysis, but only in the inner 3 pc ($\sim10'$) and based on only 4 bins.

In Fig. \ref{fig:pm_radial_distr} we show the distribution of $\mu_r$ for the stars in the inner 18$'$. 
The median (orange vertical line) is found at $0.11 \pm 0.02\,$mas\,yr$^{-1} $ = $0.52 \pm 0.08\,$ km\, s$^{-1}$, indicating an expansion. Previously, \citet{Kuhn:berk59_data_young_clusters_gaia} derived a value of $0.34 \pm 0.24\,$km\, s$^{-1}$, or $0.33 \pm 0.22\,$km\,s$^{-1}$ if rescaled to the distance of 1009\,pc. Due to the large uncertainties, \citet{Kuhn:berk59_data_young_clusters_gaia} conclude that the results on expansion/contraction for Berkeley 59 are ambiguous. Our member sample, which is several times larger than that used in \citet{Kuhn:berk59_data_young_clusters_gaia}, yields a result with $>6\sigma$ significance in favor of cluster expansion. 
The distribution is also asymmetric, with a small secondary peak to the right of the median, a hint of which can also be seen in Fig. 5.1 of \citet{Kuhn:berk59_data_young_clusters_gaia}.

\section{Properties of Berkeley 59}
\label{sec:be59}

In this section, we concentrate on the central cluster of the region, Berkeley 59, limiting the analysis to the $18'$ radius around the central position, as determined in Section~\ref{sec:centre_rad_distr}.

\subsection{Masses and ages from the HR diagram}
\label{sec:masses}
We determine stellar ages and masses using the HR diagram (Fig.~\ref{fig:hrd}). To achieve this, we interpolate PARSEC isochrones for ages ranging from 0.1 Myr to 
100 Myr in steps of 0.05 dex. To estimate the uncertainties in the derived parameters, we conduct a Monte Carlo simulation in which $T_{\mathrm{eff}}$ and M$_J$ are re-sampled based on their uncertainties. In the SED fittig procedure (Section~\ref{sec:sed_hr}), we assigned the uncertainties in $T_{\mathrm{eff}}$ and A$_V$ equal to the step in the fitting grid (100\,K and 0.25\,mag, respectively). However, $T_{\mathrm{eff}}$ and A$_V$ from the SED fitting appear to be correlated, as shown in Appendix~\ref{sec:teff_av_corr}. We quantify this correlation using the global Pearson correlation coefficient computed across the full dataset. This coefficient is then used to construct a bivariate normal distribution for each star, centered on its derived ($T_{\mathrm{eff}}$, A$_V$) values and incorporating the corresponding uncertainties. We draw 100 random samples per star from this distribution, and independently sample the apparent $J-$band magnitudes using their reported errors. With this, we can calculate a sample of 100 M$_J$ values for each star.
The uncertainty in distance is not considered in this calculation. 

The Monte Carlo simulation described above results in mass and age distributions for each object. These are often non-Gaussian and can be significantly asymmetric. For mass, we store the derived distributions and later use them to generate random samples for deriving the IMF. For age, we record the median values from 
100 Monte Carlo iterations for each object.
In total, we derived masses for 920 objects out of the 1030 high-probability candidate members. The remaining 110 objects either do not have a $J-$band measurement or show a poor SED fit.

As is commonly seen in HR diagrams of star-forming regions,
objects in Berkeley show a significant span in ages, with the mean and median ages of 4.6\,Myr and 2.9\,Myr, respectively. The result is similar if we consider the entire studied region, or if we restrict ourselves to the 18$'$ radius around the center. The median age is in agreement with the age of $\sim2\,$Myr typically assumed for Berkeley 59 \citep{pandey2008}.

\subsection{Initial mass function}
\label{sec:imf}

\begin{figure}
  \resizebox{\hsize}{!}{\includegraphics{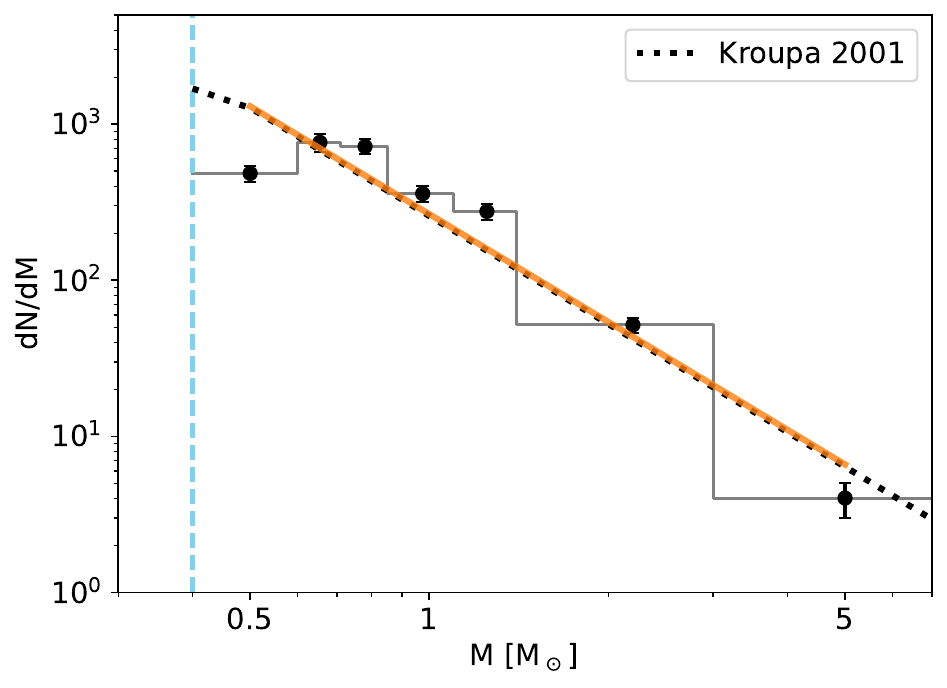}}
  \caption{ Initial mass function of Berkeley~59.
   The solid orange line shows a power law fit
   with a slope $\alpha$ = 2.3$\pm$0.2. The standard mass function from \citet{kroupa01} normalized to match the total number of objects in our IMF derivation is shown by the dotted black line. 
 The vertical blue dashed line marks the completeness limit.}
  \label{fig:imf}
\end{figure}

The IMF has been derived using the procedure identical to that from \citet{Muzic-etal:random_forest_rosette}, where the masses for each star are randomly sampled from the respective mass distributions obtained in Sect.~\ref{sec:masses}, with additional bootstrapping. This results in 10000 mass distributions, which are binned onto the same grid and then averaged. The bins have been selected as to contain a significant number of stars (typically 80-100), except from the highest-mass bin which is more sparsely populated ($\sim$20 stars).

The IMF is derived for the mass range between 0.4\,M$_\odot$ and 7\,M$_\odot$. The low-mass limit is set by the estimated completeness limit (Section~\ref{sec:completeness}), while the high-mass limit stems from the upper $T_{\mathrm{eff}}$ limit in the SED fitting (Section~\ref{sec:sed_hr}). In Appendix~\ref{app:selfunc}, we examine the potential impact of the selection steps on the IMF derivation and conclude that, within the chosen mass range, the selection process is not expected to introduce significant biases.

In Fig.~\ref{fig:imf}, we show the derived IMF. The vertical dashed blue line marks the completeness limit at 0.4\,M$_{\odot}$. With the dotted black line, we show the standard IMF from \citet{kroupa01}, which is identical to the \citet{salpeter55} IMF down to 0.5\,M$_{\odot}$. The orange line is the fit to the data, which is in excellent agreement with the standard IMF. A power-law fit in the form $dN/dM\propto M^{-\alpha}$ yields the slope $\alpha$ = $2.3 \pm 0.3$. A similar slope ($\alpha$ = $2.33 \pm 0.11$.) has been previously derived for the same cluster, for the mass range between 0.2 and 28\,M$_{\odot}$ by \citet{Panwar:Berkeley59_2018}

\subsection{Velocity dispersion and relaxation timescale}
\label{sec:vel_disp}

One-dimensional velocity dispersion $\sigma_{1D}$ is obtained by combining in quadrature the velocity dispersions in right ascension and declination (See Eq. 3 in \citet{Muzic-etal:random_forest_rosette}). 
These in turn are determined as the weighted standard deviations of the measured proper motions in the two directions. To ensure a robust velocity dispersion measurement, the proper motions are resampled $10^4$ times, incorporating proper motion uncertainties, their correlations, and bootstrapping. This yields $\sigma_{1D}$ = 0.25$\pm$0.03 mas\,yr$^{-1}$, equivalent to 1.18$\pm$0.12 km\,s$^{-1}$. This value is in agreement with  $\sigma_{1D}$=1.2$\pm$0.2 km\,s$^{-1}$ derived in \citet{Kuhn:berk59_data_young_clusters_gaia}.

The total mass of Berkeley 59 (out to the radius of 18$'$) and half-mass radius can be estimated using an MC simulation similar to the one used to derive the IMF. We obtain the total mass of 650$\pm$30\,M$_{\odot}$, which is in principle a lower limit, due to the completeness at the lower-mass end, and the lack of sensitivity of the optical data to the more obscured YSO population. The half-mass radius r$_h$= $5.8\pm0.4'$, equivalent to $1.71\pm0.13$\,pc at the distance of 1009\,pc. Using the mass and the half-mass radius, we can calculate the virial velocity dispersion $\sigma_{vir}$ (Eq. 4 in \citealt{Muzic-etal:random_forest_rosette}). Assuming $\eta$=10, we obtain $\sigma_{vir}$=$0.41\pm0.02\,$km\,s$^{-1}$. However, smaller values of $\eta$ may be appropriate for clusters with $\gamma< 3$ \citep{portegieszwart10}. For $\eta=5$, we obtain $\sigma_{vir}$=$0.57\pm0.02\,$km\,s$^{-1}$. In any case, the cluster appears to be super-virial at the present time.

Using the half-mass radius and the one-dimensional velocity dispersion, we can calculate the crossing time  \citep{binneytremaine} as $t_{\mathrm{cross}} \approx$ $r_{h}$/$\sigma_{1D}$ = $1.4\pm0.2\,$Myr. Combined with N=707 (number of probable members within r$=18'$), we derive the relaxation time (see Eq. 5 in \citealt{Muzic-etal:random_forest_rosette}) $t_{\mathrm{relax}} \approx$$15\pm2\,$Myr. The relaxation time is significantly longer than the estimated cluster age. This suggests that the mass segregation reported in \citet{Panwar:Berkeley59_2018,Panwar2024} is probably primordial and not a consequence of dynamical relaxation. 

To assess whether the cluster may be influenced by the Galactic potential, we can compare its radius with its Jacobi (or Hill) radius \citep{binneytremaine}. To that end, we use Eq. (6) in \citet{Muzic-etal:random_forest_rosette}, which requires the knowledge if the cluster mass and the Oort's constants at its Galactic position. The latter can be estimated using the relations from \citet{piskunov07}. We obtain the Jacobi radius r$_J \approx 12.5\,$pc, which is more than twice as large as the outer cluster radius we determined in Section~\ref{sec:centre_rad_distr}.

\begin{table}
\centering
\small 
\caption{Summary of various kinematic and structural parameters of Berkeley 59. }
\label{tab:parameters}
\begin{tabular}{llcc}\hline\hline
Parameter   & Value   & Value at 1009 pc & Sect.\\
\hline
$\sigma_{1D}$ &  0.25$\pm$0.03\,mas\,yr$^{-1}$  &  1.18$\pm$ 0.12\,km\,s$^{-1}$ & \ref{sec:vel_disp}\\
$\sigma_{IQR,1D}$ &  0.20$\pm$0.07\,mas\,yr$^{-1}$  &  0.99$\pm$0.35\,km\,s$^{-1}$ & \ref{sec:vel_disp} \\
$\sigma_{vir}$ &  $\sim0.1$\,mas\,yr$^{-1}$  &  $\sim 0.5$\,km\,s$^{-1}$ & \ref{sec:vel_disp}\\
M$_{tot}$ &$650\pm30$ M$_\odot$ & &  \ref{sec:vel_disp}\\
Age &  $\sim$2.9\,Myr& & \ref{sec:masses} \\
r$_{c}$ & $1.72\pm 0.25'$ & $0.50\pm 0.07$\,pc & \ref{sec:centre_rad_distr} \\
r$_h$ & $5.8\pm0.4'$ & $1.71\pm0.13$\,pc & \ref{sec:vel_disp} \\
r$_J$ & $\approx43'$ & $\approx 12.5$\,pc & \ref{sec:vel_disp}\\
t$_{cross}$ & $1.4\pm$0.2\,Myr & & \ref{sec:vel_disp}\\
t$_{relax}$ & $15\pm2$\,Myr & & \ref{sec:vel_disp}\\
$\Sigma_{0}$ &  $14.0\pm1.9$\,stars\,arcmin$^{-2}$ & 163$\pm 22$\,stars\,pc$^{-2}$ &  \ref{sec:centre_rad_distr}\\
$\Sigma_{mean}$ & $\sim$3\,stars\,arcmin$^{-2}$ & $\sim$35 stars\,pc$^{-2}$  &  \ref{sec:centre_rad_distr}\\
\hline
\end{tabular}
\end{table}

\section{Summary and conclusions}
\label{sec:summary}

In this work, we study a region with a radius of 2$\degr$ in Cepheus OB4, centered on the young cluster Berkeley~59.
Using a catalog that includes optical and near-infrared photometry, along with precise positions and proper motions from $Gaia$\,DR3, we applied the PRF algorithm to estimate membership probabilities for each source within our field of view. Stellar masses and extinctions were determined by fitting SEDs to atmospheric models, and membership was further refined by requiring a position in the HR diagram compatible with youth. Based on a list of 1030 probable members, we investigate the internal dynamics of the region relative to the mean proper motion of Berkeley~59. The main findings of this work are summarized below.

\begin{itemize}

    \item Most of the sources are concentrated within the cluster Berkeley~59 or scattered around it. Additionally, a small but distinct group of stars is found about 1$\degr$ north of Berkeley 59, associated with the nebula NGC 7822 and located at a similar distance. We also identify a number of potentially young sources south of Berkeley~59, which roughly follow the filamentary structure of the molecular gas but appear to be situated significantly behind Cepheus OB4. 

    \item The distance to Berkeley~59 estimated from Gaia\,DR3 parallaxes is $1009 \pm 12\,$pc, and the median age is 2.9\,Myr, in agreement with previous age estimates.

     \item   The radial profile of Berkeley~59 was fitted with the EFF profile \citep{Elson:radial_profile}, which returns the peak stellar surface density $\Sigma_0$ = $14\pm1.9$\,stars\,arcmin$^{-2}$ , and the (King profile) core radius r$_c$ = $0.50 \pm 0.07\,$pc. We estimate that the cluster extends out to a radius of $\sim 18\,$arcmin, which is equivalent to $\sim$5\,pc at a distance of 1009\,pc.

      \item   Berkeley~59 is showing an expansion pattern, with an expansion velocity that increases with radius. The median of the distribution of the radial component of the relative proper motion vector ($\mu_r$) out to r$=18'$ is at $0.11\pm0.02\,$mas\,yr$^{-1}$, which is equivalent to v$_r$ = $0.52 \pm 0.08$\,km\,s$^{-1}$ at a distance of 1009\,pc, i.e. the expansion is confirmed with a statistical significance of $>6\sigma$. Moreover, the detected expansion pattern is asymmetric, with the preferred direction towards the north. This aligns with the findings of \citet{wright24}, who report that most expanding young clusters and OB associations exhibit asymmetric expansion.

     \item  The IMF between 0.4 and 7\,M$_\odot$, is well represented by a single power law ($dN/dM\propto M^{-\alpha}$), with the slope $\alpha = 2.3 \pm 0.3$, close to Salpeter's slope and in agreement with previous works in the same region.

     \item The velocity dispersion of Berkeley~59 is well above the virial velocity dispersion derived from the total mass ($650\pm30$\,M$_\odot$) and half-mass radius ($1.71\pm0.13$\,pc). 
      
      \item The relaxation timescale is several times larger than the estimated age of Berkeley~59, suggesting that the cluster is still not dynamically relaxed. 

     \item The proper motions of the sources in the nebula NGC 7822 (Cluster $\#0$ in \citealt{mintz21}) point away from Berkeley~59. This region also contains a larger fraction of Class I sources than Berkeley~59 \citep{mintz21}, suggesting that it may have been triggered by expansion of the HII region.        
\end{itemize}

\section*{Data availability}
Table~\ref{tab:members} is only available in electronic form at the CDS via anonymous ftp to cdsarc.u-strasbg.fr (130.79.128.5) or via http://cdsweb.u-strasbg.fr/cgi-bin/qcat?J/A+A/.

\begin{acknowledgements} 
We thank Santiago González Gaitán and the anonymous referee for valuable discussions and insightful suggestions that improved the quality of this work. K.M. acknowledges support from the Fundação para a Ciência e a Tecnologia (FCT) through the CEEC-individual contract 2022.03809.CEECIND and research grants UIDB/04434/2020 and UIDP/04434/2020, and the Scientific Visitor Programme of the European Southern Observatory (ESO) in Chile. V.A-A acknowledges support from the INAF grant 1.05.12.05.03.
This work has made use of data from the European Space Agency (ESA) mission
{\it Gaia} (\url{https://www.cosmos.esa.int/gaia}), processed by the {\it Gaia}
Data Processing and Analysis Consortium (DPAC,
\url{https://www.cosmos.esa.int/web/gaia/dpac/consortium}). Funding for the DPAC
has been provided by national institutions, in particular the institutions
participating in the {\it Gaia} Multilateral Agreement. This publication makes use of VOSA, developed under the Spanish Virtual Observatory (https://svo.cab.inta-csic.es) project funded by MCIN/AEI/10.13039/501100011033/ through grant PID2020-112949GB-I00.
VOSA has been partially updated by using funding from the European Union's Horizon 2020 Research and Innovation Programme, under Grant Agreement nº 776403 (EXOPLANETS-A).
\end{acknowledgements}

%Literature

\bibliographystyle{aa} % style aa.bst
\bibliography{literature.bib} % your references Yourfile.bib

\begin{thebibliography}{71}
\expandafter\ifx\csname natexlab\endcsname\relax\def\natexlab#1{#1}\fi

\bibitem[{{Allard} {et~al.}(2011){Allard}, {Homeier}, \& {Freytag}}]{allard11}
{Allard}, F., {Homeier}, D., \& {Freytag}, B. 2011, in Astronomical Society of
  the Pacific Conference Series, Vol. 448, 16th Cambridge Workshop on Cool
  Stars, Stellar Systems, and the Sun, ed. C.~{Johns-Krull}, M.~K. {Browning},
  \& A.~A. {West}, 91

\bibitem[{{Armstrong} \& {Tan}(2024)}]{armstrong24}
{Armstrong}, J.~J. \& {Tan}, J.~C. 2024, \aap, 692, A166

\bibitem[{{Armstrong} {et~al.}(2022){Armstrong}, {Wright}, {Jeffries},
  {Jackson}, \& {Cantat-Gaudin}}]{armstrong22}
{Armstrong}, J.~J., {Wright}, N.~J., {Jeffries}, R.~D., {Jackson}, R.~J., \&
  {Cantat-Gaudin}, T. 2022, \mnras, 517, 5704

\bibitem[{{Bailer-Jones}(2011)}]{bailer-jones11}
{Bailer-Jones}, C.~A.~L. 2011, \mnras, 411, 435

\bibitem[{{Baraffe} {et~al.}(2015){Baraffe}, {Homeier}, {Allard}, \&
  {Chabrier}}]{baraffe15}
{Baraffe}, I., {Homeier}, D., {Allard}, F., \& {Chabrier}, G. 2015, \aap, 577,
  A42

\bibitem[{{Bayo} {et~al.}(2017){Bayo}, {Barrado}, {Allard}, {Henning},
  {Comer{\'o}n}, {Morales-Calder{\'o}n}, {Rajpurohit}, {Pe{\~n}a Ram{\'\i}rez},
  \& {Beam{\'\i}n}}]{bayo17}
{Bayo}, A., {Barrado}, D., {Allard}, F., {et~al.} 2017, \mnras, 465, 760

\bibitem[{{Binney} \& {Tremaine}(2008)}]{binneytremaine}
{Binney}, J. \& {Tremaine}, S. 2008, {Galactic Dynamics: Second Edition}
  (Princeton, N.J. : Princeton University Press)

\bibitem[{{Bressan} {et~al.}(2012){Bressan}, {Marigo}, {Girardi}, {Salasnich},
  {Dal Cero}, {Rubele}, \& {Nanni}}]{Bressan-etal:parsec_isochrone}
{Bressan}, A., {Marigo}, P., {Girardi}, L., {et~al.} 2012, \mnras, 427, 127

\bibitem[{{Cantat-Gaudin} {et~al.}(2018){Cantat-Gaudin}, {Jordi}, {Vallenari},
  {Bragaglia}, {Balaguer-N{\'u}{\~n}ez}, {Soubiran}, {Bossini}, {Moitinho},
  {Castro-Ginard}, {Krone-Martins}, {Casamiquela}, {Sordo}, \&
  {Carrera}}]{Cantat-etal:max_likelihood_distance}
{Cantat-Gaudin}, T., {Jordi}, C., {Vallenari}, A., {et~al.} 2018, \aap, 618,
  A93

\bibitem[{{Cantat-Gaudin} {et~al.}(2019){Cantat-Gaudin}, {Jordi}, {Wright},
  {Armstrong}, {Vallenari}, {Balaguer-N{\'u}{\~n}ez}, {Ramos}, {Bossini},
  {Padoan}, {Pelkonen}, {Mapelli}, \& {Jeffries}}]{cantat-gaudin19}
{Cantat-Gaudin}, T., {Jordi}, C., {Wright}, N.~J., {et~al.} 2019, \aap, 626,
  A17

\bibitem[{{Conrad} {et~al.}(2017){Conrad}, {Scholz}, {Kharchenko}, {Piskunov},
  {R{\"o}ser}, {Schilbach}, {de Jong}, {Schnurr}, {Steinmetz}, {Grebel},
  {Zwitter}, {Bienaym{\'e}}, {Bland-Hawthorn}, {Gibson}, {Gilmore},
  {Kordopatis}, {Kunder}, {Navarro}, {Parker}, {Reid}, {Seabroke}, {Siviero},
  {Watson}, \& {Wyse}}]{conrad2017}
{Conrad}, C., {Scholz}, R.~D., {Kharchenko}, N.~V., {et~al.} 2017, \aap, 600,
  A106

\bibitem[{{Della Croce} {et~al.}(2024){Della Croce}, {Dalessandro},
  {Livernois}, \& {Vesperini}}]{dellacroce24}
{Della Croce}, A., {Dalessandro}, E., {Livernois}, A., \& {Vesperini}, E. 2024,
  \aap, 683, A10

\bibitem[{{Elson} {et~al.}(1987){Elson}, {Fall}, \&
  {Freeman}}]{Elson:radial_profile}
{Elson}, R. A.~W., {Fall}, S.~M., \& {Freeman}, K.~C. 1987, \apj, 323, 54

\bibitem[{{Eswaraiah} {et~al.}(2012){Eswaraiah}, {Pandey}, {Maheswar}, {Chen},
  {Ojha}, \& {Chandola}}]{eswaraiah12}
{Eswaraiah}, C., {Pandey}, A.~K., {Maheswar}, G., {et~al.} 2012, \mnras, 419,
  2587

\bibitem[{{Flewelling}(2017)}]{panstarrs:data_release}
{Flewelling}, H. 2017, in American Astronomical Society Meeting Abstracts, Vol.
  229, American Astronomical Society Meeting Abstracts \#229, 237.07

\bibitem[{{Gahm} {et~al.}(2022){Gahm}, {Wilhelm}, {Persson}, {Djupvik}, \&
  {Portegies Zwart}}]{gahm2022}
{Gahm}, G.~F., {Wilhelm}, M.~J.~C., {Persson}, C.~M., {Djupvik}, A.~A., \&
  {Portegies Zwart}, S.~F. 2022, \aap, 663, A111

\bibitem[{{Gaia Collaboration} {et~al.}(2018){Gaia Collaboration}, {Helmi},
  {van Leeuwen}, {McMillan}, {Massari}, {Antoja}, {Robin}, {Lindegren},
  {Bastian}, {Arenou}, {Babusiaux}, {Biermann}, {Breddels}, {Hobbs}, {Jordi},
  {Pancino}, {Reyl{\'e}}, {Veljanoski}, {Brown}, {Vallenari}, {Prusti}, {de
  Bruijne}, {Bailer-Jones}, {Evans}, {Eyer}, {Jansen}, {Klioner}, {Lammers},
  {Luri}, {Mignard}, {Panem}, {Pourbaix}, {Randich}, {Sartoretti}, {Siddiqui},
  {Soubiran}, {Walton}, {Cropper}, {Drimmel}, {Katz}, {Lattanzi}, {Bakker},
  {Cacciari}, {Casta{\~n}eda}, {Chaoul}, {Cheek}, {De Angeli}, {Fabricius},
  {Guerra}, {Holl}, {Masana}, {Messineo}, {Mowlavi}, {Nienartowicz}, {Panuzzo},
  {Portell}, {Riello}, {Seabroke}, {Tanga}, {Th{\'e}venin}, {Gracia-Abril},
  {Comoretto}, {Garcia-Reinaldos}, {Teyssier}, {Altmann}, {Andrae}, {Audard},
  {Bellas-Velidis}, {Benson}, {Berthier}, {Blomme}, {Burgess}, {Busso},
  {Carry}, {Cellino}, {Clementini}, {Clotet}, {Creevey}, {Davidson}, {De
  Ridder}, {Delchambre}, {Dell'Oro}, {Ducourant},
  {Fern{\'a}ndez-Hern{\'a}ndez}, {Fouesneau}, {Fr{\'e}mat}, {Galluccio},
  {Garc{\'\i}a-Torres}, {Gonz{\'a}lez-N{\'u}{\~n}ez}, {Gonz{\'a}lez-Vidal},
  {Gosset}, {Guy}, {Halbwachs}, {Hambly}, {Harrison}, {Hern{\'a}ndez},
  {Hestroffer}, {Hodgkin}, {Hutton}, {Jasniewicz}, {Jean-Antoine-Piccolo},
  {Jordan}, {Korn}, {Krone-Martins}, {Lanzafame}, {Lebzelter}, {L{\"o}ffler},
  {Manteiga}, {Marrese}, {Mart{\'\i}n-Fleitas}, {Moitinho}, {Mora}, {Muinonen},
  {Osinde}, {Pauwels}, {Petit}, {Recio-Blanco}, {Richards}, {Rimoldini},
  {Sarro}, {Siopis}, {Smith}, {Sozzetti}, {S{\"u}veges}, {Torra}, {van Reeven},
  {Abbas}, {Abreu Aramburu}, {Accart}, {Aerts}, {Altavilla}, {{\'A}lvarez},
  {Alvarez}, {Alves}, {Anderson}, {Andrei}, {Anglada Varela}, {Antiche},
  {Arcay}, {Astraatmadja}, {Bach}, {Baker}, {Balaguer-N{\'u}{\~n}ez}, {Balm},
  {Barache}, {Barata}, {Barbato}, {Barblan}, {Barklem}, {Barrado}, {Barros},
  {Barstow}, {Bartholom{\'e} Mu{\~n}oz}, {Bassilana}, {Becciani}, {Bellazzini},
  {Berihuete}, {Bertone}, {Bianchi}, {Bienaym{\'e}}, {Blanco-Cuaresma}, {Boch},
  {Boeche}, {Bombrun}, {Borrachero}, {Bossini}, {Bouquillon}, {Bourda},
  {Bragaglia}, {Bramante}, {Bressan}, {Brouillet}, {Br{\"u}semeister},
  {Brugaletta}, {Bucciarelli}, {Burlacu}, {Busonero}, {Butkevich}, {Buzzi},
  {Caffau}, {Cancelliere}, {Cannizzaro}, {Cantat-Gaudin}, {Carballo},
  {Carlucci}, {Carrasco}, {Casamiquela}, {Castellani}, {Castro-Ginard},
  {Charlot}, {Chemin}, {Chiavassa}, {Cocozza}, {Costigan}, {Cowell}, {Crifo},
  {Crosta}, {Crowley}, {Cuypers}, {Dafonte}, {Damerdji}, {Dapergolas}, {David},
  {David}, {de Laverny}, {De Luise}, {De March}, {de Martino}, {de Souza}, {de
  Torres}, {Debosscher}, {del Pozo}, {Delbo}, {Delgado}, {Delgado}, {Di
  Matteo}, {Diakite}, {Diener}, {Distefano}, {Dolding}, {Drazinos},
  {Dur{\'a}n}, {Edvardsson}, {Enke}, {Eriksson}, {Esquej}, {Eynard Bontemps},
  {Fabre}, {Fabrizio}, {Faigler}, {Falc{\~a}o}, {Farr{\`a}s Casas}, {Federici},
  {Fedorets}, {Fernique}, {Figueras}, {Filippi}, {Findeisen}, {Fonti},
  {Fraile}, {Fraser}, {Fr{\'e}zouls}, {Gai}, {Galleti}, {Garabato},
  {Garc{\'\i}a-Sedano}, {Garofalo}, {Garralda}, {Gavel}, {Gavras}, {Gerssen},
  {Geyer}, {Giacobbe}, {Gilmore}, {Girona}, {Giuffrida}, {Glass}, {Gomes},
  {Granvik}, {Gueguen}, {Guerrier}, {Guiraud}, {Guti{\'e}rrez-S{\'a}nchez},
  {Hofmann}, {Holland}, {Huckle}, {Hypki}, {Icardi}, {Jan{\ss}en}, {Jevardat de
  Fombelle}, {Jonker}, {Juh{\'a}sz}, {Julbe}, {Karampelas}, {Kewley}, {Klar},
  {Kochoska}, {Kohley}, {Kolenberg}, {Kontizas}, {Kontizas}, {Koposov},
  {Kordopatis}, {Kostrzewa-Rutkowska}, {Koubsky}, {Lambert}, {Lanza}, {Lasne},
  {Lavigne}, {Le Fustec}, {Le Poncin-Lafitte}, {Lebreton}, {Leccia}, {Leclerc},
  {Lecoeur-Taibi}, {Lenhardt}, {Leroux}, {Liao}, {Licata}, {Lindstr{\o}m},
  {Lister}, {Livanou}, {Lobel}, {L{\'o}pez}, {Managau}, {Mann}, {Mantelet},
  {Marchal}, {Marchant}, {Marconi}, {Marinoni}, {Marschalk{\'o}}, {Marshall},
  {Martino}, {Marton}, {Mary}, {Matijevi{\v{c}}}, {Mazeh}, {Messina},
  {Michalik}, {Millar}, {Molina}, {Molinaro}, {Moln{\'a}r}, {Montegriffo},
  {Mor}, {Morbidelli}, {Morel}, {Morris}, {Mulone}, {Muraveva}, {Musella},
  {Nelemans}, {Nicastro}, {Noval}, {O'Mullane}, {Ord{\'e}novic},
  {Ord{\'o}{\~n}ez-Blanco}, {Osborne}, {Pagani}, {Pagano}, {Pailler},
  {Palacin}, {Palaversa}, {Panahi}, {Pawlak}, {Piersimoni}, {Pineau}, {Plachy},
  {Plum}, {Poggio}, {Poujoulet}, {Pr{\v{s}}a}, {Pulone}, {Racero}, {Ragaini},
  {Rambaux}, {Ramos-Lerate}, {Regibo}, {Riclet}, {Ripepi}, {Riva}, {Rivard},
  {Rixon}, {Roegiers}, {Roelens}, {Romero-G{\'o}mez}, {Rowell}, {Royer},
  {Ruiz-Dern}, {Sadowski}, {Sagrist{\`a} Sell{\'e}s}, {Sahlmann}, {Salgado},
  {Salguero}, {Sanna}, {Santana-Ros}, {Sarasso}, {Savietto}, {Schultheis},
  {Sciacca}, {Segol}, {Segovia}, {S{\'e}gransan}, {Shih}, {Siltala}, {Silva},
  {Smart}, {Smith}, {Solano}, {Solitro}, {Sordo}, {Soria Nieto}, {Souchay},
  {Spagna}, {Spoto}, {Stampa}, {Steele}, {Steidelm{\"u}ller}, {Stephenson},
  {Stoev}, {Suess}, {Surdej}, {Szabados}, {Szegedi-Elek}, {Tapiador}, {Taris},
  {Tauran}, {Taylor}, {Teixeira}, {Terrett}, {Teyssandier}, {Thuillot},
  {Titarenko}, {Torra Clotet}, {Turon}, {Ulla}, {Utrilla}, {Uzzi}, {Vaillant},
  {Valentini}, {Valette}, {van Elteren}, {Van Hemelryck}, {van Leeuwen},
  {Vaschetto}, {Vecchiato}, {Viala}, {Vicente}, {Vogt}, {von Essen}, {Voss},
  {Votruba}, {Voutsinas}, {Walmsley}, {Weiler}, {Wertz}, {Wevems},
  {Wyrzykowski}, {Yoldas}, {{\v{Z}}erjal}, {Ziaeepour}, {Zorec}, {Zschocke},
  {Zucker}, {Zurbach}, \& {Zwitter}}]{gaiacollab2018:persp_correction}
{Gaia Collaboration}, {Helmi}, A., {van Leeuwen}, F., {et~al.} 2018, \aap, 616,
  A12

\bibitem[{{Gaia Collaboration} {et~al.}(2016){Gaia Collaboration}, {Prusti},
  {de Bruijne}, {Brown}, {Vallenari}, {Babusiaux}, {Bailer-Jones}, {Bastian},
  {Biermann}, {Evans}, {Eyer}, {Jansen}, {Jordi}, {Klioner}, {Lammers},
  {Lindegren}, {Luri}, {Mignard}, {Milligan}, {Panem}, {Poinsignon},
  {Pourbaix}, {Randich}, {Sarri}, {Sartoretti}, {Siddiqui}, {Soubiran},
  {Valette}, {van Leeuwen}, {Walton}, {Aerts}, {Arenou}, {Cropper}, {Drimmel},
  {H{\o}g}, {Katz}, {Lattanzi}, {O'Mullane}, {Grebel}, {Holland}, {Huc},
  {Passot}, {Bramante}, {Cacciari}, {Casta{\~n}eda}, {Chaoul}, {Cheek}, {De
  Angeli}, {Fabricius}, {Guerra}, {Hern{\'a}ndez}, {Jean-Antoine-Piccolo},
  {Masana}, {Messineo}, {Mowlavi}, {Nienartowicz}, {Ord{\'o}{\~n}ez-Blanco},
  {Panuzzo}, {Portell}, {Richards}, {Riello}, {Seabroke}, {Tanga},
  {Th{\'e}venin}, {Torra}, {Els}, {Gracia-Abril}, {Comoretto},
  {Garcia-Reinaldos}, {Lock}, {Mercier}, {Altmann}, {Andrae}, {Astraatmadja},
  {Bellas-Velidis}, {Benson}, {Berthier}, {Blomme}, {Busso}, {Carry},
  {Cellino}, {Clementini}, {Cowell}, {Creevey}, {Cuypers}, {Davidson}, {De
  Ridder}, {de Torres}, {Delchambre}, {Dell'Oro}, {Ducourant}, {Fr{\'e}mat},
  {Garc{\'\i}a-Torres}, {Gosset}, {Halbwachs}, {Hambly}, {Harrison}, {Hauser},
  {Hestroffer}, {Hodgkin}, {Huckle}, {Hutton}, {Jasniewicz}, {Jordan},
  {Kontizas}, {Korn}, {Lanzafame}, {Manteiga}, {Moitinho}, {Muinonen},
  {Osinde}, {Pancino}, {Pauwels}, {Petit}, {Recio-Blanco}, {Robin}, {Sarro},
  {Siopis}, {Smith}, {Smith}, {Sozzetti}, {Thuillot}, {van Reeven}, {Viala},
  {Abbas}, {Abreu Aramburu}, {Accart}, {Aguado}, {Allan}, {Allasia},
  {Altavilla}, {{\'A}lvarez}, {Alves}, {Anderson}, {Andrei}, {Anglada Varela},
  {Antiche}, {Antoja}, {Ant{\'o}n}, {Arcay}, {Atzei}, {Ayache}, {Bach},
  {Baker}, {Balaguer-N{\'u}{\~n}ez}, {Barache}, {Barata}, {Barbier}, {Barblan},
  {Baroni}, {Barrado y Navascu{\'e}s}, {Barros}, {Barstow}, {Becciani},
  {Bellazzini}, {Bellei}, {Bello Garc{\'\i}a}, {Belokurov}, {Bendjoya},
  {Berihuete}, {Bianchi}, {Bienaym{\'e}}, {Billebaud}, {Blagorodnova},
  {Blanco-Cuaresma}, {Boch}, {Bombrun}, {Borrachero}, {Bouquillon}, {Bourda},
  {Bouy}, {Bragaglia}, {Breddels}, {Brouillet}, {Br{\"u}semeister},
  {Bucciarelli}, {Budnik}, {Burgess}, {Burgon}, {Burlacu}, {Busonero}, {Buzzi},
  {Caffau}, {Cambras}, {Campbell}, {Cancelliere}, {Cantat-Gaudin}, {Carlucci},
  {Carrasco}, {Castellani}, {Charlot}, {Charnas}, {Charvet}, {Chassat},
  {Chiavassa}, {Clotet}, {Cocozza}, {Collins}, {Collins}, {Costigan}, {Crifo},
  {Cross}, {Crosta}, {Crowley}, {Dafonte}, {Damerdji}, {Dapergolas}, {David},
  {David}, {De Cat}, {de Felice}, {de Laverny}, {De Luise}, {De March}, {de
  Martino}, {de Souza}, {Debosscher}, {del Pozo}, {Delbo}, {Delgado},
  {Delgado}, {di Marco}, {Di Matteo}, {Diakite}, {Distefano}, {Dolding}, {Dos
  Anjos}, {Drazinos}, {Dur{\'a}n}, {Dzigan}, {Ecale}, {Edvardsson}, {Enke},
  {Erdmann}, {Escolar}, {Espina}, {Evans}, {Eynard Bontemps}, {Fabre},
  {Fabrizio}, {Faigler}, {Falc{\~a}o}, {Farr{\`a}s Casas}, {Faye}, {Federici},
  {Fedorets}, {Fern{\'a}ndez-Hern{\'a}ndez}, {Fernique}, {Fienga}, {Figueras},
  {Filippi}, {Findeisen}, {Fonti}, {Fouesneau}, {Fraile}, {Fraser}, {Fuchs},
  {Furnell}, {Gai}, {Galleti}, {Galluccio}, {Garabato}, {Garc{\'\i}a-Sedano},
  {Gar{\'e}}, {Garofalo}, {Garralda}, {Gavras}, {Gerssen}, {Geyer}, {Gilmore},
  {Girona}, {Giuffrida}, {Gomes}, {Gonz{\'a}lez-Marcos},
  {Gonz{\'a}lez-N{\'u}{\~n}ez}, {Gonz{\'a}lez-Vidal}, {Granvik}, {Guerrier},
  {Guillout}, {Guiraud}, {G{\'u}rpide}, {Guti{\'e}rrez-S{\'a}nchez}, {Guy},
  {Haigron}, {Hatzidimitriou}, {Haywood}, {Heiter}, {Helmi}, {Hobbs},
  {Hofmann}, {Holl}, {Holland}, {Hunt}, {Hypki}, {Icardi}, {Irwin}, {Jevardat
  de Fombelle}, {Jofr{\'e}}, {Jonker}, {Jorissen}, {Julbe}, {Karampelas},
  {Kochoska}, {Kohley}, {Kolenberg}, {Kontizas}, {Koposov}, {Kordopatis},
  {Koubsky}, {Kowalczyk}, {Krone-Martins}, {Kudryashova}, {Kull}, {Bachchan},
  {Lacoste-Seris}, {Lanza}, {Lavigne}, {Le Poncin-Lafitte}, {Lebreton},
  {Lebzelter}, {Leccia}, {Leclerc}, {Lecoeur-Taibi}, {Lemaitre}, {Lenhardt},
  {Leroux}, {Liao}, {Licata}, {Lindstr{\o}m}, {Lister}, {Livanou}, {Lobel},
  {L{\"o}ffler}, {L{\'o}pez}, {Lopez-Lozano}, {Lorenz}, {Loureiro},
  {MacDonald}, {Magalh{\~a}es Fernandes}, {Managau}, {Mann}, {Mantelet},
  {Marchal}, {Marchant}, {Marconi}, {Marie}, {Marinoni}, {Marrese},
  {Marschalk{\'o}}, {Marshall}, {Mart{\'\i}n-Fleitas}, {Martino}, {Mary},
  {Matijevi{\v{c}}}, {Mazeh}, {McMillan}, {Messina}, {Mestre}, {Michalik},
  {Millar}, {Miranda}, {Molina}, {Molinaro}, {Molinaro}, {Moln{\'a}r},
  {Moniez}, {Montegriffo}, {Monteiro}, {Mor}, {Mora}, {Morbidelli}, {Morel},
  {Morgenthaler}, {Morley}, {Morris}, {Mulone}, {Muraveva}, {Musella},
  {Narbonne}, {Nelemans}, {Nicastro}, {Noval}, {Ord{\'e}novic},
  {Ordieres-Mer{\'e}}, {Osborne}, {Pagani}, {Pagano}, {Pailler}, {Palacin},
  {Palaversa}, {Parsons}, {Paulsen}, {Pecoraro}, {Pedrosa}, {Pentik{\"a}inen},
  {Pereira}, {Pichon}, {Piersimoni}, {Pineau}, {Plachy}, {Plum}, {Poujoulet},
  {Pr{\v{s}}a}, {Pulone}, {Ragaini}, {Rago}, {Rambaux}, {Ramos-Lerate},
  {Ranalli}, {Rauw}, {Read}, {Regibo}, {Renk}, {Reyl{\'e}}, {Ribeiro},
  {Rimoldini}, {Ripepi}, {Riva}, {Rixon}, {Roelens}, {Romero-G{\'o}mez},
  {Rowell}, {Royer}, {Rudolph}, {Ruiz-Dern}, {Sadowski}, {Sagrist{\`a}
  Sell{\'e}s}, {Sahlmann}, {Salgado}, {Salguero}, {Sarasso}, {Savietto},
  {Schnorhk}, {Schultheis}, {Sciacca}, {Segol}, {Segovia}, {Segransan},
  {Serpell}, {Shih}, {Smareglia}, {Smart}, {Smith}, {Solano}, {Solitro},
  {Sordo}, {Soria Nieto}, {Souchay}, {Spagna}, {Spoto}, {Stampa}, {Steele},
  {Steidelm{\"u}ller}, {Stephenson}, {Stoev}, {Suess}, {S{\"u}veges}, {Surdej},
  {Szabados}, {Szegedi-Elek}, {Tapiador}, {Taris}, {Tauran}, {Taylor},
  {Teixeira}, {Terrett}, {Tingley}, {Trager}, {Turon}, {Ulla}, {Utrilla},
  {Valentini}, {van Elteren}, {Van Hemelryck}, {van Leeuwen}, {Varadi},
  {Vecchiato}, {Veljanoski}, {Via}, {Vicente}, {Vogt}, {Voss}, {Votruba},
  {Voutsinas}, {Walmsley}, {Weiler}, {Weingrill}, {Werner}, {Wevers},
  {Whitehead}, {Wyrzykowski}, {Yoldas}, {{\v{Z}}erjal}, {Zucker}, {Zurbach},
  {Zwitter}, {Alecu}, {Allen}, {Allende Prieto}, {Amorim},
  {Anglada-Escud{\'e}}, {Arsenijevic}, {Azaz}, {Balm}, {Beck}, {Bernstein},
  {Bigot}, {Bijaoui}, {Blasco}, {Bonfigli}, {Bono}, {Boudreault}, {Bressan},
  {Brown}, {Brunet}, {Bunclark}, {Buonanno}, {Butkevich}, {Carret}, {Carrion},
  {Chemin}, {Ch{\'e}reau}, {Corcione}, {Darmigny}, {de Boer}, {de Teodoro}, {de
  Zeeuw}, {Delle Luche}, {Domingues}, {Dubath}, {Fodor}, {Fr{\'e}zouls},
  {Fries}, {Fustes}, {Fyfe}, {Gallardo}, {Gallegos}, {Gardiol}, {Gebran},
  {Gomboc}, {G{\'o}mez}, {Grux}, {Gueguen}, {Heyrovsky}, {Hoar}, {Iannicola},
  {Isasi Parache}, {Janotto}, {Joliet}, {Jonckheere}, {Keil}, {Kim},
  {Klagyivik}, {Klar}, {Knude}, {Kochukhov}, {Kolka}, {Kos}, {Kutka}, {Lainey},
  {LeBouquin}, {Liu}, {Loreggia}, {Makarov}, {Marseille}, {Martayan},
  {Martinez-Rubi}, {Massart}, {Meynadier}, {Mignot}, {Munari}, {Nguyen},
  {Nordlander}, {Ocvirk}, {O'Flaherty}, {Olias Sanz}, {Ortiz}, {Osorio},
  {Oszkiewicz}, {Ouzounis}, {Palmer}, {Park}, {Pasquato}, {Peltzer}, {Peralta},
  {P{\'e}turaud}, {Pieniluoma}, {Pigozzi}, {Poels}, {Prat}, {Prod'homme},
  {Raison}, {Rebordao}, {Risquez}, {Rocca-Volmerange}, {Rosen}, {Ruiz-Fuertes},
  {Russo}, {Sembay}, {Serraller Vizcaino}, {Short}, {Siebert}, {Silva},
  {Sinachopoulos}, {Slezak}, {Soffel}, {Sosnowska}, {Strai{\v{z}}ys}, {ter
  Linden}, {Terrell}, {Theil}, {Tiede}, {Troisi}, {Tsalmantza}, {Tur},
  {Vaccari}, {Vachier}, {Valles}, {Van Hamme}, {Veltz}, {Virtanen}, {Wallut},
  {Wichmann}, {Wilkinson}, {Ziaeepour}, \&
  {Zschocke}}]{gaiamission:gaiacollab:2016}
{Gaia Collaboration}, {Prusti}, T., {de Bruijne}, J.~H.~J., {et~al.} 2016,
  \aap, 595, A1

\bibitem[{{Gaia Collaboration} {et~al.}(2023){Gaia Collaboration}, {Vallenari},
  {Brown}, {Prusti}, {de Bruijne}, {Arenou}, {Babusiaux}, {Biermann},
  {Creevey}, {Ducourant}, {Evans}, {Eyer}, {Guerra}, {Hutton}, {Jordi},
  {Klioner}, {Lammers}, {Lindegren}, {Luri}, {Mignard}, {Panem}, {Pourbaix},
  {Randich}, {Sartoretti}, {Soubiran}, {Tanga}, {Walton}, {Bailer-Jones},
  {Bastian}, {Drimmel}, {Jansen}, {Katz}, {Lattanzi}, {van Leeuwen}, {Bakker},
  {Cacciari}, {Casta{\~n}eda}, {De Angeli}, {Fabricius}, {Fouesneau},
  {Fr{\'e}mat}, {Galluccio}, {Guerrier}, {Heiter}, {Masana}, {Messineo},
  {Mowlavi}, {Nicolas}, {Nienartowicz}, {Pailler}, {Panuzzo}, {Riclet}, {Roux},
  {Seabroke}, {Sordo}, {Th{\'e}venin}, {Gracia-Abril}, {Portell}, {Teyssier},
  {Altmann}, {Andrae}, {Audard}, {Bellas-Velidis}, {Benson}, {Berthier},
  {Blomme}, {Burgess}, {Busonero}, {Busso}, {C{\'a}novas}, {Carry}, {Cellino},
  {Cheek}, {Clementini}, {Damerdji}, {Davidson}, {de Teodoro}, {Nu{\~n}ez
  Campos}, {Delchambre}, {Dell'Oro}, {Esquej}, {Fern{\'a}ndez-Hern{\'a}ndez},
  {Fraile}, {Garabato}, {Garc{\'\i}a-Lario}, {Gosset}, {Haigron}, {Halbwachs},
  {Hambly}, {Harrison}, {Hern{\'a}ndez}, {Hestroffer}, {Hodgkin}, {Holl},
  {Jan{\ss}en}, {Jevardat de Fombelle}, {Jordan}, {Krone-Martins}, {Lanzafame},
  {L{\"o}ffler}, {Marchal}, {Marrese}, {Moitinho}, {Muinonen}, {Osborne},
  {Pancino}, {Pauwels}, {Recio-Blanco}, {Reyl{\'e}}, {Riello}, {Rimoldini},
  {Roegiers}, {Rybizki}, {Sarro}, {Siopis}, {Smith}, {Sozzetti}, {Utrilla},
  {van Leeuwen}, {Abbas}, {{\'A}brah{\'a}m}, {Abreu Aramburu}, {Aerts},
  {Aguado}, {Ajaj}, {Aldea-Montero}, {Altavilla}, {{\'A}lvarez}, {Alves},
  {Anders}, {Anderson}, {Anglada Varela}, {Antoja}, {Baines}, {Baker},
  {Balaguer-N{\'u}{\~n}ez}, {Balbinot}, {Balog}, {Barache}, {Barbato},
  {Barros}, {Barstow}, {Bartolom{\'e}}, {Bassilana}, {Bauchet}, {Becciani},
  {Bellazzini}, {Berihuete}, {Bernet}, {Bertone}, {Bianchi}, {Binnenfeld},
  {Blanco-Cuaresma}, {Blazere}, {Boch}, {Bombrun}, {Bossini}, {Bouquillon},
  {Bragaglia}, {Bramante}, {Breedt}, {Bressan}, {Brouillet}, {Brugaletta},
  {Bucciarelli}, {Burlacu}, {Butkevich}, {Buzzi}, {Caffau}, {Cancelliere},
  {Cantat-Gaudin}, {Carballo}, {Carlucci}, {Carnerero}, {Carrasco},
  {Casamiquela}, {Castellani}, {Castro-Ginard}, {Chaoul}, {Charlot}, {Chemin},
  {Chiaramida}, {Chiavassa}, {Chornay}, {Comoretto}, {Contursi}, {Cooper},
  {Cornez}, {Cowell}, {Crifo}, {Cropper}, {Crosta}, {Crowley}, {Dafonte},
  {Dapergolas}, {David}, {David}, {de Laverny}, {De Luise}, \& {De
  March}}]{gaiad3:gaiacollab:2022}
{Gaia Collaboration}, {Vallenari}, A., {Brown}, A.~G.~A., {et~al.} 2023, \aap,
  674, A1

\bibitem[{Getman {et~al.}(2017)Getman, Broos, Kuhn, Feigelson, Richert, Ota,
  Bate, \& Garmire}]{Getman-etal:SFinCs}
Getman, K.~V., Broos, P.~S., Kuhn, M.~A., {et~al.} 2017, The Astrophysical
  Journal Supplement Series, 229, 28

\bibitem[{{Goodwin} \& {Bastian}(2006)}]{goodwin2006}
{Goodwin}, S.~P. \& {Bastian}, N. 2006, \mnras, 373, 752

\bibitem[{{Guilherme-Garcia} {et~al.}(2023){Guilherme-Garcia}, {Krone-Martins},
  \& {Moitinho}}]{guilherme-garcia23}
{Guilherme-Garcia}, P., {Krone-Martins}, A., \& {Moitinho}, A. 2023, \aap, 673,
  A128

\bibitem[{{Gutermuth} {et~al.}(2009){Gutermuth}, {Megeath}, {Myers}, {Allen},
  {Pipher}, \& {Fazio}}]{gutermuth09}
{Gutermuth}, R.~A., {Megeath}, S.~T., {Myers}, P.~C., {et~al.} 2009, \apjs,
  184, 18

\bibitem[{{Inutsuka} {et~al.}(2015){Inutsuka}, {Inoue}, {Iwasaki}, \&
  {Hosokawa}}]{inutsuka15}
{Inutsuka}, S.-i., {Inoue}, T., {Iwasaki}, K., \& {Hosokawa}, T. 2015, \aap,
  580, A49

\bibitem[{{Kharchenko} {et~al.}(2007){Kharchenko}, {Scholz}, {Piskunov},
  {R{\"o}ser}, \& {Schilbach}}]{kharchenko2009}
{Kharchenko}, N.~V., {Scholz}, R.~D., {Piskunov}, A.~E., {R{\"o}ser}, S., \&
  {Schilbach}, E. 2007, Astronomische Nachrichten, 328, 889

\bibitem[{{King}(1966)}]{king66}
{King}, I.~R. 1966, \aj, 71, 276

\bibitem[{{Koenig} {et~al.}(2012){Koenig}, {Leisawitz}, {Benford}, {Rebull},
  {Padgett}, \& {Assef}}]{koenig12}
{Koenig}, X.~P., {Leisawitz}, D.~T., {Benford}, D.~J., {et~al.} 2012, \apj,
  744, 130

\bibitem[{{Kroupa}(2001)}]{kroupa01}
{Kroupa}, P. 2001, \mnras, 322, 231

\bibitem[{{Kuhn} {et~al.}(2014){Kuhn}, {Feigelson}, {Getman}, {Baddeley},
  {Broos}, {Sills}, {Bate}, {Povich}, {Luhman}, {Busk}, {Naylor}, \&
  {King}}]{kuhn14}
{Kuhn}, M.~A., {Feigelson}, E.~D., {Getman}, K.~V., {et~al.} 2014, \apj, 787,
  107

\bibitem[{{Kuhn} {et~al.}(2017){Kuhn}, {Getman}, {Feigelson}, {Sills},
  {Gromadzki}, {Medina}, {Borissova}, \& {Kurtev}}]{kuhn17}
{Kuhn}, M.~A., {Getman}, K.~V., {Feigelson}, E.~D., {et~al.} 2017, \aj, 154,
  214

\bibitem[{{Kuhn} {et~al.}(2019){Kuhn}, {Hillenbrand}, {Sills}, {Feigelson}, \&
  {Getman}}]{Kuhn:berk59_data_young_clusters_gaia}
{Kuhn}, M.~A., {Hillenbrand}, L.~A., {Sills}, A., {Feigelson}, E.~D., \&
  {Getman}, K.~V. 2019, \apj, 870, 32

\bibitem[{{Kun} {et~al.}(2008){Kun}, {Kiss}, \& {Balog}}]{kun2008}
{Kun}, M., {Kiss}, Z.~T., \& {Balog}, Z. 2008, in Handbook of Star Forming
  Regions, Volume I, ed. B.~{Reipurth}, Vol.~4, 136

\bibitem[{{Lada} \& {Lada}(2003)}]{lada&lada03}
{Lada}, C.~J. \& {Lada}, E.~A. 2003, \araa, 41, 57

\bibitem[{{Lemaitre} {et~al.}(2016){Lemaitre}, {Nogueira}, \&
  {Aridas}}]{Lemaitre-NAC:imbalanced_learn}
{Lemaitre}, G., {Nogueira}, F., \& {Aridas}, C.~K. 2016, arXiv e-prints,
  arXiv:1609.06570

\bibitem[{{Lim} {et~al.}(2021){Lim}, {Naz{\'e}}, {Hong}, {Park}, {Yun}, {Yi},
  {Park}, {Hwang}, \& {Lee}}]{lim21}
{Lim}, B., {Naz{\'e}}, Y., {Hong}, J., {et~al.} 2021, \aj, 162, 56

\bibitem[{{Lindegren} {et~al.}(2021){Lindegren}, {Bastian}, {Biermann},
  {Bombrun}, {de Torres}, {Gerlach}, {Geyer}, {Hern{\'a}ndez}, {Hilger},
  {Hobbs}, {Klioner}, {Lammers}, {McMillan}, {Ramos-Lerate},
  {Steidelm{\"u}ller}, {Stephenson}, \& {van
  Leeuwen}}]{Lindegren-etal:gaia_bias}
{Lindegren}, L., {Bastian}, U., {Biermann}, M., {et~al.} 2021, \aap, 649, A4

\bibitem[{{Lozinskaya} {et~al.}(1987){Lozinskaya}, {Sitnik}, \&
  {Toropova}}]{lozinskaya87}
{Lozinskaya}, T.~A., {Sitnik}, T.~G., \& {Toropova}, M.~S. 1987, \sovast, 31,
  493

\bibitem[{{Majaess} {et~al.}(2008){Majaess}, {Turner}, {Lane}, \&
  {Moncrieff}}]{majaess08}
{Majaess}, D.~J., {Turner}, D.~G., {Lane}, D.~J., \& {Moncrieff}, K.~E. 2008,
  The Journal of the American Association of Variable Star Observers, 36, 90

\bibitem[{Matthews(1975)}]{mcc:matthew}
Matthews, B. 1975, Biochimica et Biophysica Acta (BBA) - Protein Structure,
  405, 442

\bibitem[{{Meingast} {et~al.}(2021){Meingast}, {Alves}, \&
  {Rottensteiner}}]{meingast21}
{Meingast}, S., {Alves}, J., \& {Rottensteiner}, A. 2021, \aap, 645, A84

\bibitem[{{Mintz} {et~al.}(2021){Mintz}, {Hora}, \& {Winston}}]{mintz21}
{Mintz}, A., {Hora}, J.~L., \& {Winston}, E. 2021, \aj, 162, 236

\bibitem[{{Miret-Roig} {et~al.}(2019){Miret-Roig}, {Bouy}, {Olivares}, {Sarro},
  {Tamura}, {Allen}, {Bertin}, {Serre}, {Berihuete}, {Beletsky}, {Barrado},
  {Hu{\'e}lamo}, {Cuillandre}, {Moraux}, \& {Bouvier}}]{miret19}
{Miret-Roig}, N., {Bouy}, H., {Olivares}, J., {et~al.} 2019, \aap, 631, A57

\bibitem[{{Mu{\v{z}}i{\'c}} {et~al.}(2022){Mu{\v{z}}i{\'c}}, {Almendros-Abad},
  {Bouy}, {Kubiak}, {Pe{\~n}a Ram{\'\i}rez}, {Krone-Martins}, {Moitinho}, \&
  {Concei{\c{c}}{\~a}o}}]{Muzic-etal:random_forest_rosette}
{Mu{\v{z}}i{\'c}}, K., {Almendros-Abad}, V., {Bouy}, H., {et~al.} 2022, \aap,
  668, A19

\bibitem[{{Mu{\v{z}}i{\'c}} {et~al.}(2019){Mu{\v{z}}i{\'c}}, {Scholz},
  {Pe{\~n}a Ram{\'\i}rez}, {Jayawardhana}, {Sch{\"o}del}, {Geers}, {Cieza}, \&
  {Bayo}}]{Muzic-etal:deep_look_rosette}
{Mu{\v{z}}i{\'c}}, K., {Scholz}, A., {Pe{\~n}a Ram{\'\i}rez}, K., {et~al.}
  2019, \apj, 881, 79

\bibitem[{{Ogura} {et~al.}(2002){Ogura}, {Sugitani}, \& {Pickles}}]{ogura2002}
{Ogura}, K., {Sugitani}, K., \& {Pickles}, A. 2002, \aj, 123, 2597

\bibitem[{{Pandey} {et~al.}(2008){Pandey}, {Sharma}, {Ogura}, {Ojha}, {Chen},
  {Bhatt}, \& {Ghosh}}]{pandey2008}
{Pandey}, A.~K., {Sharma}, S., {Ogura}, K., {et~al.} 2008, \mnras, 383, 1241

\bibitem[{{Pang} {et~al.}(2021){Pang}, {Li}, {Yu}, {Tang}, {Dinnbier},
  {Kroupa}, {Pasquato}, \& {Kouwenhoven}}]{pang21}
{Pang}, X., {Li}, Y., {Yu}, Z., {et~al.} 2021, \apj, 912, 162

\bibitem[{{Panwar} {et~al.}(2018){Panwar}, {Pandey}, {Samal}, {Battinelli},
  {Ogura}, {Ojha}, {Chen}, \& {Singh}}]{Panwar:Berkeley59_2018}
{Panwar}, N., {Pandey}, A.~K., {Samal}, M.~R., {et~al.} 2018, \aj, 155, 44

\bibitem[{{Panwar} {et~al.}(2024){Panwar}, {Rishi}, {Sharma}, {Ojha}, {Samal},
  {Singh}, \& {Kesh Yadav}}]{Panwar2024}
{Panwar}, N., {Rishi}, C., {Sharma}, S., {et~al.} 2024, arXiv e-prints,
  arXiv:2406.08261

\bibitem[{{Parker} {et~al.}(2014){Parker}, {Wright}, {Goodwin}, \&
  {Meyer}}]{parker14}
{Parker}, R.~J., {Wright}, N.~J., {Goodwin}, S.~P., \& {Meyer}, M.~R. 2014,
  \mnras, 438, 620

\bibitem[{{Pastorelli} {et~al.}(2020){Pastorelli}, {Marigo}, {Girardi},
  {Aringer}, {Chen}, {Rubele}, {Trabucchi}, {Bladh}, {Boyer}, {Bressan},
  {Dalcanton}, {Groenewegen}, {Lebzelter}, {Mowlavi}, {Chubb}, {Cioni}, {de
  Grijs}, {Ivanov}, {Nanni}, {van Loon}, \&
  {Zaggia}}]{Pastorelli-etal:parsec_isochrone}
{Pastorelli}, G., {Marigo}, P., {Girardi}, L., {et~al.} 2020, \mnras, 498, 3283

\bibitem[{{Piskunov} {et~al.}(2007){Piskunov}, {Schilbach}, {Kharchenko},
  {R{\"o}ser}, \& {Scholz}}]{piskunov07}
{Piskunov}, A.~E., {Schilbach}, E., {Kharchenko}, N.~V., {R{\"o}ser}, S., \&
  {Scholz}, R.~D. 2007, \aap, 468, 151

\bibitem[{{Planck Collaboration}(2016)}]{Planck2016}
{Planck Collaboration}. 2016, AAP, 594, A11

\bibitem[{{Portegies Zwart} {et~al.}(2010){Portegies Zwart}, {McMillan}, \&
  {Gieles}}]{portegieszwart10}
{Portegies Zwart}, S.~F., {McMillan}, S. L.~W., \& {Gieles}, M. 2010, \araa,
  48, 431

\bibitem[{{Reis} {et~al.}(2019){Reis}, {Baron}, \& {Shahaf}}]{Reis:PRF}
{Reis}, I., {Baron}, D., \& {Shahaf}, S. 2019, \aj, 157, 16

\bibitem[{{Rossano} {et~al.}(1983){Rossano}, {Grayzeck}, \&
  {Angerhofer}}]{rossano83}
{Rossano}, G.~S., {Grayzeck}, E.~J., \& {Angerhofer}, P.~E. 1983, \aj, 88, 1835

\bibitem[{{Rosvick} \& {Majaess}(2013)}]{rosvick13}
{Rosvick}, J.~M. \& {Majaess}, D. 2013, \aj, 146, 142

\bibitem[{{Salpeter}(1955)}]{salpeter55}
{Salpeter}, E.~E. 1955, \apj, 121, 161

\bibitem[{{Sills} {et~al.}(2018){Sills}, {Rieder}, {Scora}, {McCloskey}, \&
  {Jaffa}}]{sills18}
{Sills}, A., {Rieder}, S., {Scora}, J., {McCloskey}, J., \& {Jaffa}, S. 2018,
  \mnras, 477, 1903

\bibitem[{{Skiff}(2014)}]{skiff14}
{Skiff}, B.~A. 2014, {VizieR Online Data Catalog: Catalogue of Stellar Spectral
  Classifications (Skiff, 2009-2014)}, VizieR On-line Data Catalog: B/mk.
  Originally published in: 2014yCat....1.2023S

\bibitem[{{Skrutskie} {et~al.}(2006){Skrutskie}, {Cutri}, {Stiening},
  {Weinberg}, {Schneider}, {Carpenter}, {Beichman}, {Capps}, {Chester},
  {Elias}, {Huchra}, {Liebert}, {Lonsdale}, {Monet}, {Price}, {Seitzer},
  {Jarrett}, {Kirkpatrick}, {Gizis}, {Howard}, {Evans}, {Fowler}, {Fullmer},
  {Hurt}, {Light}, {Kopan}, {Marsh}, {McCallon}, {Tam}, {Van Dyk}, \&
  {Wheelock}}]{2masscollab:2mass_data}
{Skrutskie}, M.~F., {Cutri}, R.~M., {Stiening}, R., {et~al.} 2006, \aj, 131,
  1163

\bibitem[{{Sugitani} {et~al.}(1991){Sugitani}, {Fukui}, \&
  {Ogura}}]{sugitani91}
{Sugitani}, K., {Fukui}, Y., \& {Ogura}, K. 1991, \apjs, 77, 59

\bibitem[{{Tan}(2000)}]{tan2000}
{Tan}, J.~C. 2000, \apj, 536, 173

\bibitem[{{Taylor}(2005)}]{topcat:taylor:2005}
{Taylor}, M.~B. 2005, in Astronomical Society of the Pacific Conference Series,
  Vol. 347, Astronomical Data Analysis Software and Systems XIV, ed.
  P.~{Shopbell}, M.~{Britton}, \& R.~{Ebert}, 29

\bibitem[{{van Leeuwen}(2009)}]{vanLeeuwen:persp_correction}
{van Leeuwen}, F. 2009, \aap, 497, 209

\bibitem[{{Wenger} {et~al.}(2000){Wenger}, {Ochsenbein}, {Egret}, {Dubois},
  {Bonnarel}, {Borde}, {Genova}, {Jasniewicz}, {Lalo{\"e}}, {Lesteven}, \&
  {Monier}}]{simbad:wenger:2000}
{Wenger}, M., {Ochsenbein}, F., {Egret}, D., {et~al.} 2000, \aaps, 143, 9

\bibitem[{{Wright}(2020)}]{wright2020}
{Wright}, N.~J. 2020, \nar, 90, 101549

\bibitem[{{Wright} {et~al.}(2019){Wright}, {Jeffries}, {Jackson}, {Bayo},
  {Bonito}, {Damiani}, {Kalari}, {Lanzafame}, {Pancino}, {Parker},
  {Prisinzano}, {Randich}, {Vink}, {Alfaro}, {Bergemann}, {Franciosini},
  {Gilmore}, {Gonneau}, {Hourihane}, {Jofr{\'e}}, {Koposov}, {Lewis},
  {Magrini}, {Micela}, {Morbidelli}, {Sacco}, {Worley}, \& {Zaggia}}]{wright19}
{Wright}, N.~J., {Jeffries}, R.~D., {Jackson}, R.~J., {et~al.} 2019, \mnras,
  486, 2477

\bibitem[{{Wright} {et~al.}(2024){Wright}, {Jeffries}, {Jackson}, {Sacco},
  {Arnold}, {Franciosini}, {Gilmore}, {Gonneau}, {Morbidelli}, {Prisinzano},
  {Randich}, \& {Worley}}]{wright24}
{Wright}, N.~J., {Jeffries}, R.~D., {Jackson}, R.~J., {et~al.} 2024, \mnras,
  533, 705

\bibitem[{{Yu} {et~al.}(2023){Yu}, {Khanna}, {Themessl}, {Hekker}, {Dr{\'e}au},
  {Gizon}, \& {Bi}}]{yu23}
{Yu}, J., {Khanna}, S., {Themessl}, N., {et~al.} 2023, \apjs, 264, 41

\bibitem[{{Zari} {et~al.}(2019){Zari}, {Brown}, \& {de Zeeuw}}]{zari19}
{Zari}, E., {Brown}, A.~G.~A., \& {de Zeeuw}, P.~T. 2019, \aap, 628, A123

\end{thebibliography}

\begin{appendix}
\section{PRF scores and feature importance}

\begin{figure}[h!]
 \centering
 \includegraphics[width=0.45\textwidth]{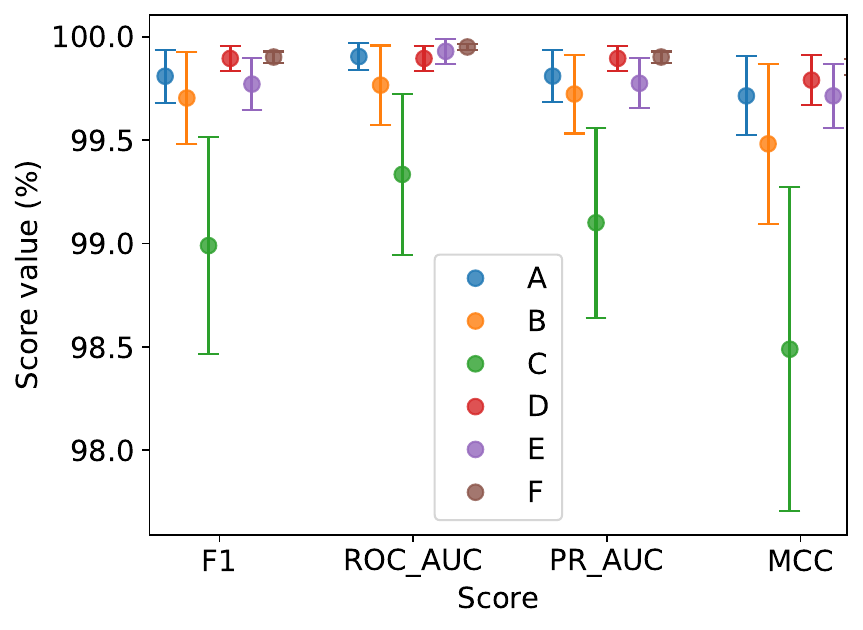}
   \caption{Metrics of the six different runs of the PRF (see Table \ref{tab:prf_sampling} for IDs, run details and the exact score values). Points are offset on the x-axis for clarity.}
  \label{fig:prf_scores}
\end{figure}

\begin{figure}[h!]
   \centering
   \includegraphics[width=0.45\textwidth]{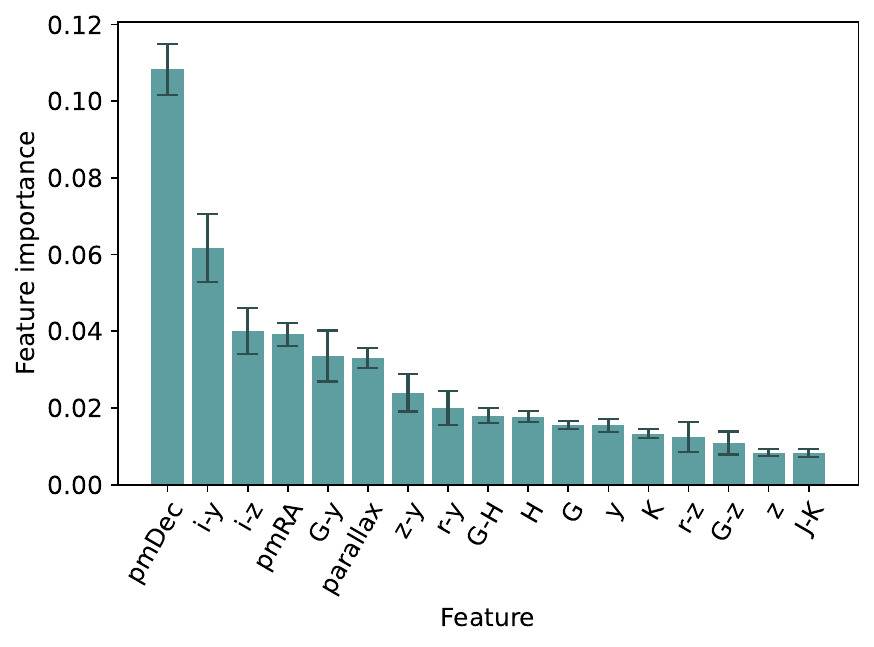}   
   \caption{Relative feature importance, evaluated for the classifier of run F. The error bars are the standard deviation of the 50 random split samples used to evaluate each classifier.}
   \label{fig:feat_imp}
\end{figure}

In Table~\ref{tab:prf_sampling} we show the details of the 6 PRF runs, along with the scores obtained
through cross-validation (see Section~\ref{sec:ev_prf}). The scores are also
shown in Fig.~\ref{fig:prf_scores}.
In Fig.~\ref{fig:feat_imp}, we show the relative importance of the 17 used features as returned by the classifier in run F. Corresponding plots from other runs look very similar to this one. The uncertainties correspond to the standard deviation of the 50 split values.

\begin{table*}
\centering
\caption{Parameters of the sampling strategy for each run of the PRF and the corresponding scores.}
		\begin{tabular}{ccccccccc}
   \hline
    \hline
			ID & under$\_$sample & over$\_$sample & N$_\text{{memb}}$ & N$_\text{{non-memb}}$ & F1($\%$) & ROC\_AUC($\%$) & PR\_AUC($\%$) & MCC($\%$) \\
			\hline
			A & 0.50 & 0.01 & 2827 & 5654 & 99.81$\pm$0.13 & 99.90$\pm$0.06 & 99.81$\pm$0.13 & 99.71$\pm$0.19 \\
			B & 0.75 & 0.005 & 1413 & 1884 & 99.70$\pm$0.22 & 99.77$\pm$0.19 & 99.72$\pm$0.19 & 99.48$\pm$0.39 \\
			C & 0.50 & 0.002 & 565 & 1130 & 98.99$\pm$0.52 & 99.33$\pm$0.39 & 99.10$\pm$0.46 & 98.49$\pm$0.78 \\
			D & 1.00 & 0.02 & 5654 & 5654 & 99.90$\pm$0.06 & 99.90$\pm$0.06 & 99.90$\pm$0.06 & 99.79$\pm$0.12 \\
			E & 0.25 & 0.01 & 2827 & 11308 & 99.77$\pm$0.12 & 99.93$\pm$0.06 & 99.78$\pm$0.12 & 99.71$\pm$0.15 \\
			F & 0.50 & 0.10 & 28271 & 56542 & 99.90$\pm$0.03 & 99.95$\pm$0.01 & 99.90$\pm$0.03 & 99.85$\pm$0.04 \\
   \hline
		\end{tabular}
	\label{tab:prf_sampling}
\end{table*}

\section{Relative Proper Motions in various regions}

\begin{figure*}[h!]
   \centering
   \includegraphics[width=17cm]{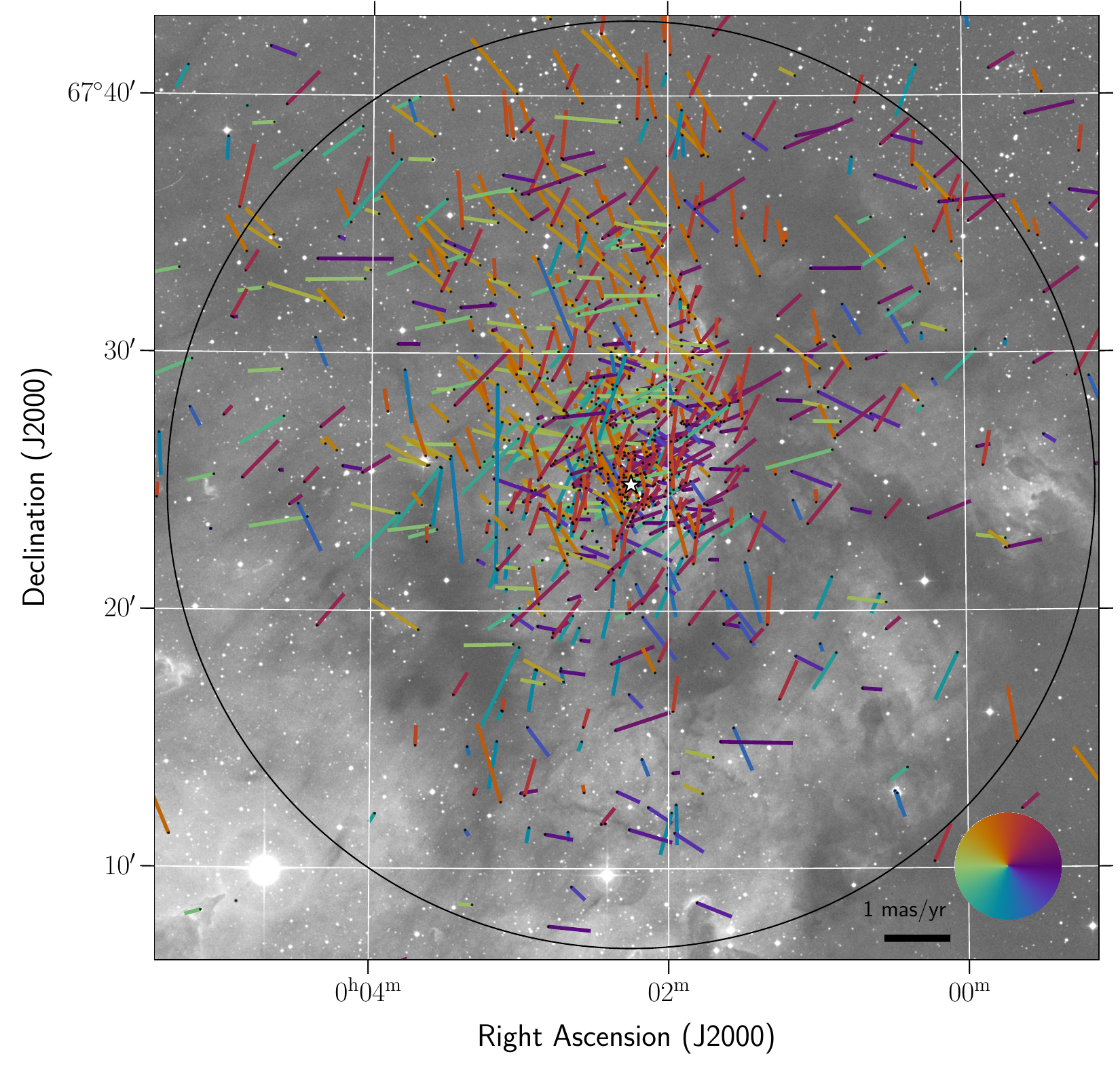}   
   \caption{Same as Fig. \ref{fig:relpm_nozoom}, but focused on the cluster Berkeley~59. The circle marks the $18'$ radius, which we adopted as the cluster radius.}
   \label{fig:relpm_zoom}
\end{figure*}

\begin{figure}[h!]
   \centering
   \includegraphics[width=0.9\linewidth]{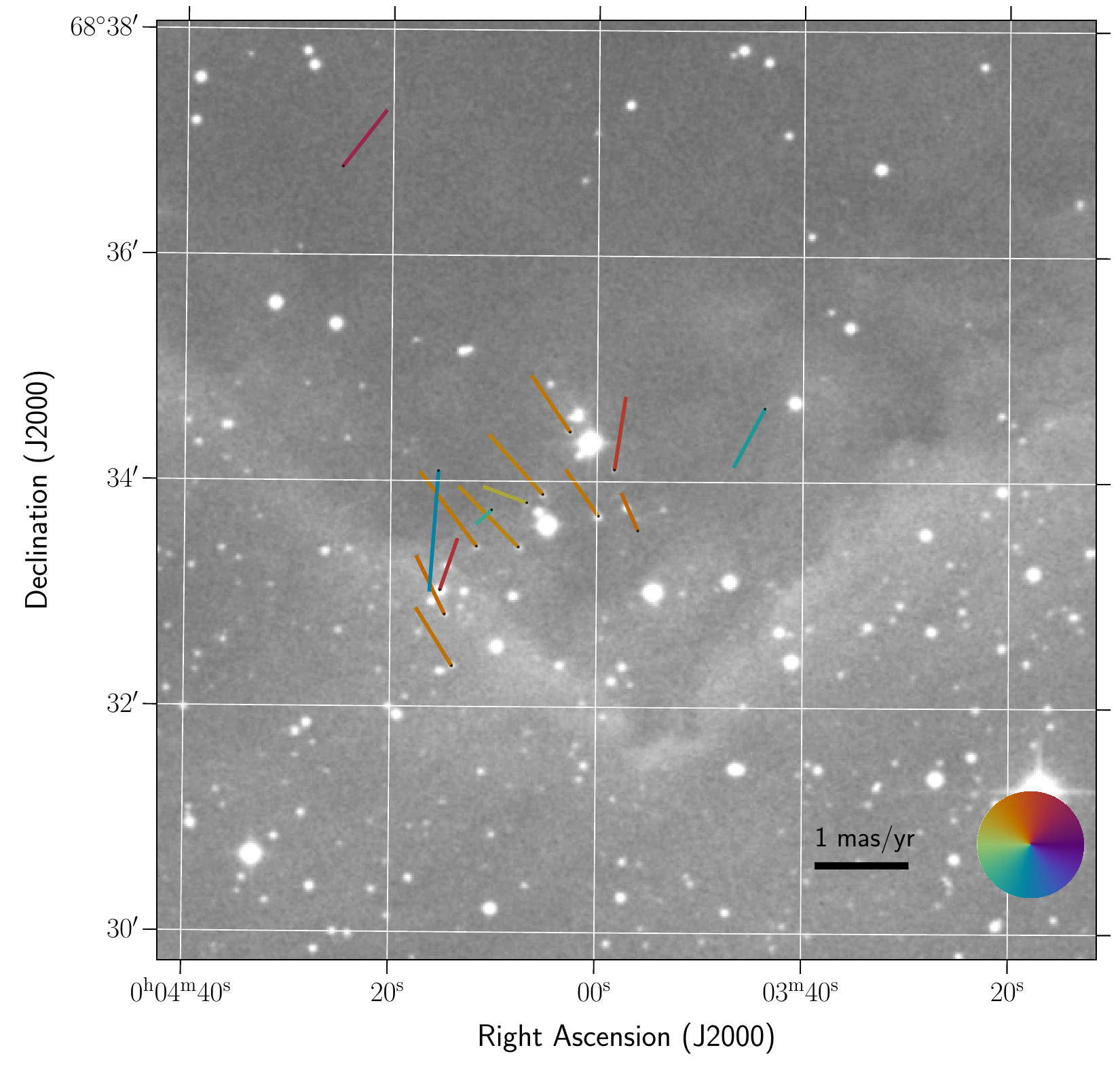}%
        \hfill
      \caption{Same as Fig.~\ref{fig:relpm_nozoom}, but focused on the small region associated with the nebula to the north of the cluster.}
         \label{fig:relpm_zoomnorth}
\end{figure}

In Figs.~\ref{fig:relpm_zoom} and \ref{fig:relpm_zoomnorth}, we show the zoom-in versions of Fig.~\ref{fig:relpm_nozoom}, which allows appreciation of details in relative proper motions for Berkeley~59 and the region $\sim 1\degr$ north of it, associated with the nebula NGC 7822 (Cluster$\#$0 in \citealt{mintz21} and BLR2 in \citealt{sugitani91, ogura2002}).

\section{Correlation between $T_{\mathrm{eff}}$ and A$_V$ from SED fitting}
\label{sec:teff_av_corr}

\begin{figure}
   \centering
   \includegraphics[width=0.95\linewidth]{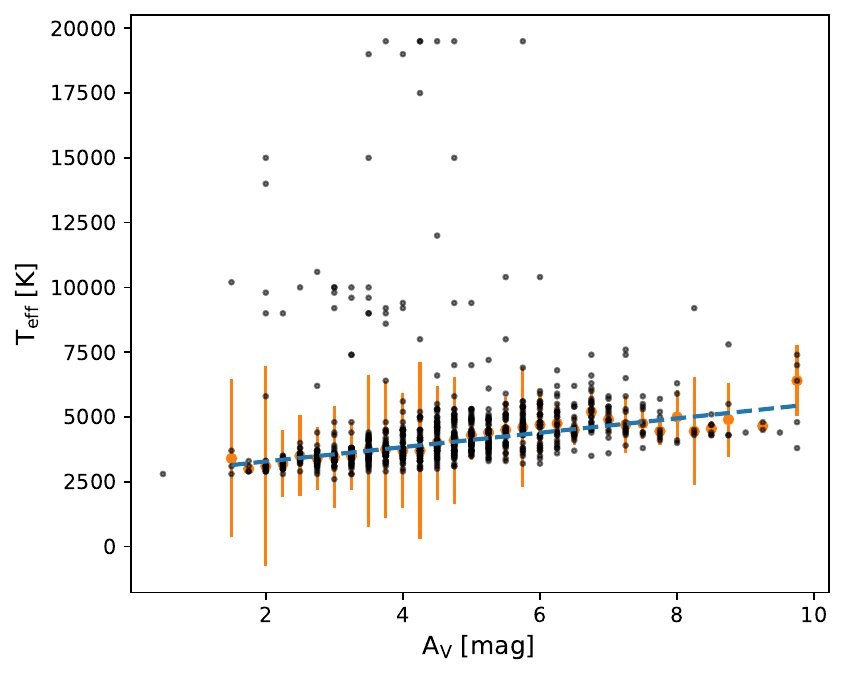}
      \caption{The relationship between the $T_{\mathrm{eff}}$ and A$_V$ obtained from the SED fitting in Section~\ref{sec:sed_hr} (black dots). The orange points represent the median value $T_{\mathrm{eff}}$ values at each step of the extinction grid, with error bars indicating the standard deviation. The blue dashed line corresponds to a linear fit to the orange points.}
         \label{fig:teff_av_corr}
\end{figure}

In Section~\ref{sec:sed_hr}, we derived $T_{\mathrm{eff}}$ and A$_V$ by fitting each star's SED to the atmosphere models. As shown in Fig~\ref{fig:teff_av_corr}, these two parameters show some degree of correlation, which is commonly seen in SED fitting using optical/near-infrared multiband photometry  
\citep{bailer-jones11, bayo17,yu23}.

\section{IMF selection function}
\label{app:selfunc}

\begin{figure}[h!]
   \centering
   \includegraphics[width=0.7\linewidth]{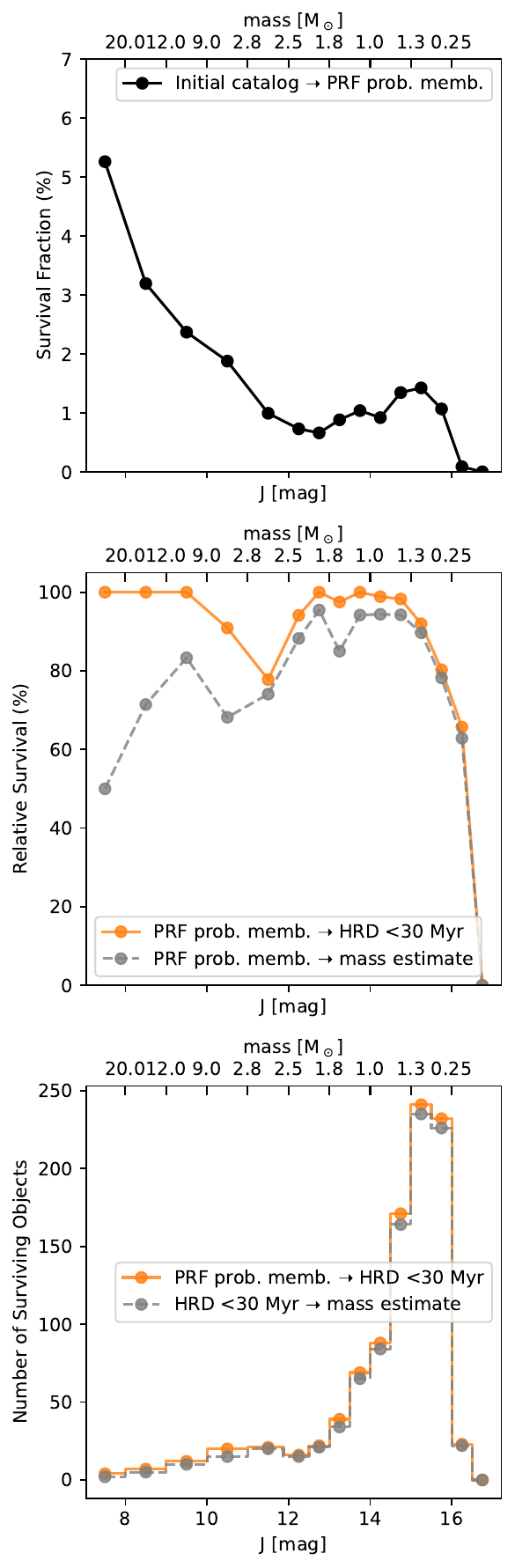}
      \caption{Selection effects as a function of $J-$band magnitude across different stages of the sample construction. For this plot, we removed the southern part of the field ($\delta < 66.8^\circ$).
Top: Fraction of objects surviving the main selection step (PRF classification).
Middle: Percentage of objects remaining at each magnitude step relative to the sample of probable members from the PRF run. The orange symbols show the percentage of surviving objects after removing the objects older than 30 Myr (Section.~\ref{sec:sed_hr}), and the grey ones those that have valid mass estimates (Section.~\ref{sec:masses}). 
Bottom: Same as the middle panel, but showing the number of objects instead of percentages. 
 Tick marks on the top x-axis indicate approximate stellar masses corresponding to $J-$band magnitudes at the distance of Berkeley 59, age of 2 Myr and extinction A$_V=4$\,mag.
      }
         \label{fig:selfunc}
\end{figure}

Figure~\ref{fig:selfunc} summarizes how the number of objects evolves through successive selection steps, shown as a function of $J-$band magnitude. For this analysis, we excluded sources with declination below $66.8^\circ$, as they are not included in the IMF calculation, which is the main focus here. We use the $J-$band magnitude because our mass estimates are derived from the HR diagram constructed in that band.

The top panel shows the fraction of sources that pass the PRF-based selection, identifying high-probability members (Section~\ref{sec:highprobcands}. The middle and right panels present statistics relative to this sample: the middle panel shows the percentage of PRF-selected sources that also meet the HRD age cut (30 Myr limit; see Section~\ref{sec:sed_hr}, orange) and that have mass estimates (grey; see Section~\ref{sec:masses}). The bottom panel shows the corresponding absolute number of these surviving sources per magnitude bin.

The decrease in the number of objects at $J\lesssim11$\,mag results from the upper limit on effective temperature during the SED fitting (Section~\ref{sec:sed_hr}). At the low-mass end, there is a sharp drop in the number of objects at $J>16$, which corresponds to masses $\lesssim$0.25\,M$_\odot$ for the distance of Berkeley\,59, age of 2 Myr, and the average extinction of A$_V=4$\,mag. For the lowest-mass bin in our IMF calculation (0.4-0.6\,M$_\odot$), the number of the removed objects is small, at most about 20$\%$. 

\end{appendix}

\end{document}